\documentclass[twocolumn,secnumarabic,amssymb, nobibnotes, aps, prd]{revtex4-1}

\newcommand{\arcsec}{\ifmmode^{\prime\prime}\else $^{\prime\prime}$\fi}
\newcommand{\arcmin}{\ifmmode^{\prime}\else $^{\prime}$\fi}
\newcommand{\degrees}{\ifmmode^{\circ}\else $^{\circ}$\fi}

\newcommand{\half}{{\textstyle{\frac{1}{2}}}}

\newcommand{\bfu}{{\bm u}} 
 \newcommand{\bfb}{{\bm b}}
 
\newcommand{\bfB}{{\bm B}}

\newcommand{\bfV}{{\bm V}} 
\newcommand{\bfP}{{\bm P}} \newcommand{\bfQ}{{\bm Q}}
  
\newcommand{\calE}{\mathcal{E}}
\newcommand{\ehalf}{\epsilon^{\frac{1}{2}}}

\newcommand{\hatzed}{\hat{\boldsymbol{z}}}

\newcommand{\av}[1]{\left\langle #1\right\rangle}

\usepackage{amssymb, amsmath}
\usepackage{graphicx}
\usepackage{bm}
\usepackage{color}
\begin{document}

\title{Kinematic dynamo action in square and hexagonal patterns}%

\author{B. Favier}%
\email[Corresponding author: ]{b.favier@damtp.cam.ac.uk}
\author{M.R.E. Proctor}
\affiliation{Department of Applied Mathematics and Theoretical Physics, University of Cambridge, Centre for Mathematical Sciences, Wilberforce Road, Cambridge CB3 0WA, UK}
\date{\today}

\begin{abstract}
We consider kinematic dynamo action in rapidly rotating Boussinesq convection just above onset.
The velocity is constrained to have either a square or a hexagonal pattern.
For the square pattern, large-scale dynamo action is observed at onset, with most of the magnetic energy being contained in the horizontally-averaged component.
As the magnetic Reynolds number increases, small-scale dynamo action becomes possible, reducing the overall growth rate of the dynamo.
For the hexagonal pattern, the breaking of symmetry between up and down flows results in an effective pumping velocity.
For intermediate rotation rates, this additional effect can prevent the growth of any mean-field dynamo, so that only a small-scale dynamo  is eventually possible at large enough magnetic Reynolds number.
For very large rotation rates, this pumping term becomes negligible, and the dynamo properties of square and hexagonal patterns are qualitatively similar.
These results hold for both perfectly conducting and infinite magnetic permeability boundary conditions.
\end{abstract}

\maketitle

%
%
\section{Introduction}

One of the principal goals of dynamo theory is to understand the origin of large-scale magnetic fields observed in stars and planets.
Although the dynamo action required certainly depends on many parameters, it is often useful to study the induction processes in simplified flows.
This approach has led to significant improvements in our understanding of the generation of magnetic fields by the motions of an electrically conducting fluid in astrophysical objects and laboratory experiments.
Kinematic dynamos driven by simplified flows are indeed very useful to model the fundamental induction mechanisms of more realistic liquid metal experiments \cite{marie2003,gissinger2009,pinter2010}.
The velocity field driving the dynamo can be measured experimentally or modelled analytically \cite{dudley89}.
The famous Roberts flow \cite{roberts70,roberts72}, whose analytical expression is very simple, is the perfect illustration of what is called a mean-field or large-scale dynamo, a mechanism which might explain the origin of magnetic fields coherent on much larger scales than the ones of the fluid motion.
In mean-field theory, which is a turbulent closure theory describing the evolution of the large-scale quantities in terms of the statistical properties of the small-scale perturbations \cite{moff78,krause80}, the evolution equation for the large-scale field is derived from the induction equation by decomposing the magnetic field into mean and fluctuating parts.
The small-scale velocity field interacts with this large-scale magnetic field creating magnetic perturbations at small scales.
Provided the flow lacks reflectional symmetry, these induced small-scale magnetic perturbations then interact with the small-scale velocity generating a non-vanishing mean electromotive force, which sustains the large-scale magnetic field.
Although this scenario is very appealing as an explanation of the existence of large-scale magnetic fields in many astrophysical objects, the situation becomes more complicated when the flow is less ideal, \textit{e.g.} turbulent, or when the magnetic Reynolds number becomes large as expected in the astrophysically relevant regime.

Of particular interest here is the well-studied topic of convectively-driven dynamos, where the flow is sustained by thermal convection between two parallel horizontal plates \citep{childress72}.
Magnetic fields of planets and stars are often accepted to be the result of convectively driven flows of an electrically conducting fluid occupying a large volume of the star or planet.
Early numerical studies have concentrated on the turbulent regime \cite{meneguzzi89,brandenburg1996,cattaneo99} where the Rayleigh number is much larger than its critical value.
However, without rotation, the flow is reflectionally-symmetric so that only small-scale dynamo action can occur.
When the plane-layer is rotating around the vertical axis, the viscous force can become of secondary importance in comparison to the Lorentz force and the flow is thus strongly controlled
by the forces exerted by the magnetic field.
Fully three-dimensional dynamo solutions in the rapid rotation limit were numerically studied by several authors \cite{jones2000,rotvig02,stell04}.
It has also been shown by \cite{cattaneo06} that turbulent moderately rotating Boussinesq convection, while breaking reflectional symmetry as required by mean-field theory, is not necessarily capable of sustaining a dynamo of mean-field type.
It is certainly able to sustain a small-scale dynamo, but the magnetic field is then locally regenerated by the stretching properties of the flow and is strongly intermittent without large-scale coherence.
Since then, several studies have tried to clarify the problem \cite{tobias2008,kapyla2009,favier2012a}, but it seems that a definitive answer is still elusive.
More recently, a transition has been shown to occur between two different types of dynamos in rapidly rotating Boussinesq convection \cite{tilgner2013}.
In order to clarify the transitions between rotationally-dominated and more turbulent dynamos, it therefore seems interesting to consider rotating convection just above onset where the flow is much more coherent spatially and temporally than in the turbulent regime.
Several studies have considered the kinematic dynamo action driven by simple patterns of convection such as rolls, squares and hexagons without rotation \cite{zheligo98,matthews99,demircan2002}.
In the rotating case, the pioneering work of \cite{soward74} has shown the existence of a large-scale dynamo in the limit of rapid rotation.

We here consider the kinematic dynamo problem in a steady velocity field corresponding to rotating Boussinesq convection just above onset.
While we neglect here the effect of the Lorentz force, our model is sufficient to derive interesting results concerning the transition between large-scale and small-scale dynamos, as well as the surprising consequence of the so-called turbulent pumping effect \cite{droby74,moff78,krause80,tobias2001}.
Near onset, the preferred pattern consists of rolls providing the system is symmetric with respect to the mid-layer.
However, at sufficiently large rotation rate, these rolls are unstable to the K\"uppers-Lortz instability \cite{kuppers1969}.
In this case, a given set rolls is unstable to another set of rolls with a different orientation.
This new set of rolls is equally unstable to yet another, leading to spatiotemporal chaos.
Surprisingly, at even larger rotation rates, square patterns were experimentally found to be stable \cite{bajaj1998}.
The existence of this square pattern, slowly rotating in the prograde direction, was later confirmed numerically by \cite{sanchez2005}.
In the case of convection lacking the up-down symmetry, either due to temperature-dependent viscosity or non-Boussinesq effects, theory predicts that the hexagonal pattern is the primary instability \citep{soward85,echebarria2000} (which is also the case without rotation).
In this paper, the cell pattern is imposed to be either squares or hexagons, for which the corresponding velocity fields are analytically known in the Boussinesq approximation \citep{veronis59}.
While we do not self-consistently solve the momentum equation in the present study, the previous discussion gives some justifications for the existence of square and hexagonal patterns in rotating plane-layer convection.
Note that such steady patterns might not be relevant to liquid metals and planetary dynamos, since the very low Prandtl number implies that the bifurcation to convection is time-dependent in that case.
However, the mechanism discussed in this paper might still be relevant to more realistic dynamos, provided that the topological magnetic pumping plays an important role.

In the next section, we describe the model and the numerical approach used to solve the induction equation with a prescribed velocity field in three dimensions.
A mean-field model is then derived and the associated results are discussed in section \ref{sec:mean}.
Sections \ref{sec:square} and \ref{sec:hexa} are devoted to the results from direct numerical simulations of the dynamo driven by square and hexagonal patterns, respectively.
Finally, we explore the effect of changing the magnetic boundary conditions in section \ref{sec:bou}.

%
%
\section{Description of the model\label{sec:model}}
We consider the evolution of a plane-parallel layer of incompressible fluid, bounded above and below by two impenetrable, stress-free walls, a distance $d$ apart.
The geometry of this layer is defined by a Cartesian grid, with $x$ and $y$ corresponding to the horizontal coordinates.
The $z$-axis points vertically downwards.
The layer is rotating about the $z$-axis, with a constant angular velocity $\bm{\Omega}=\Omega\hat{\bm{z}}$.
The horizontal size of the fluid domain is defined by the aspect ratios $\lambda_x$ and $\lambda_y$ so that the fluid occupies the domain $0<z<d$, $0<x<\lambda_xd$ and $0<y<\lambda_y d$.
The physical properties of the fluid, namely the kinematic viscosity $\nu$ and magnetic diffusivity $\eta$, are assumed to be constant.

The velocity field is imposed to be a cellular flow corresponding to the onset of Boussinesq convection in a rotating layer.
In particular, we consider the solutions first obtained by Veronis \cite{veronis59}.
We focus here on the particular cases of square and hexagonal patterns.
The velocity field associated with the square pattern is
\begin{multline}
\label{eq:square1}
u_x=-\frac{\pi}{a\sqrt{2}}\Big(\sin\frac{ax}{\sqrt2}\cos\frac{ay}{\sqrt2} \\ +\frac{\sqrt T}{\pi^2+a^2}\cos\frac{ax}{\sqrt2}\sin\frac{ay}{\sqrt2}\Big)\cos\pi z
\end{multline}
\begin{multline}
\label{eq:square2}
u_y=-\frac{\pi}{a\sqrt{2}}\Big(\cos\frac{ax}{\sqrt2}\sin\frac{ay}{\sqrt2} \\ -\frac{\sqrt T}{\pi^2+a^2}\sin\frac{ax}{\sqrt2}\cos\frac{ay}{\sqrt2}\Big)\cos\pi z
\end{multline}
\begin{equation}
\label{eq:square3}
u_z=\cos\frac{ax}{\sqrt2}\cos\frac{ay}{\sqrt2}\sin\pi z
\end{equation}
where $T$ is the Taylor number defined as $T=4\Omega^2d^4/\nu^2$ and $a$ is is the most unstable wave number in the large Taylor number limit given by \cite{chan61}
\begin{equation}
\label{eq:a}
a=\left(\frac12\pi^2T\right)^{1/6} \ .
\end{equation}
The second terms in the right-hand side of equations \eqref{eq:square1} and \eqref{eq:square2} are $O(1)$ whereas the first terms vary like $T^{-1/6}$. For large Taylor numbers, which is the focus of this paper, we therefore expect the second terms to be dominant.

\begin{figure*}
   \includegraphics[width=40mm]{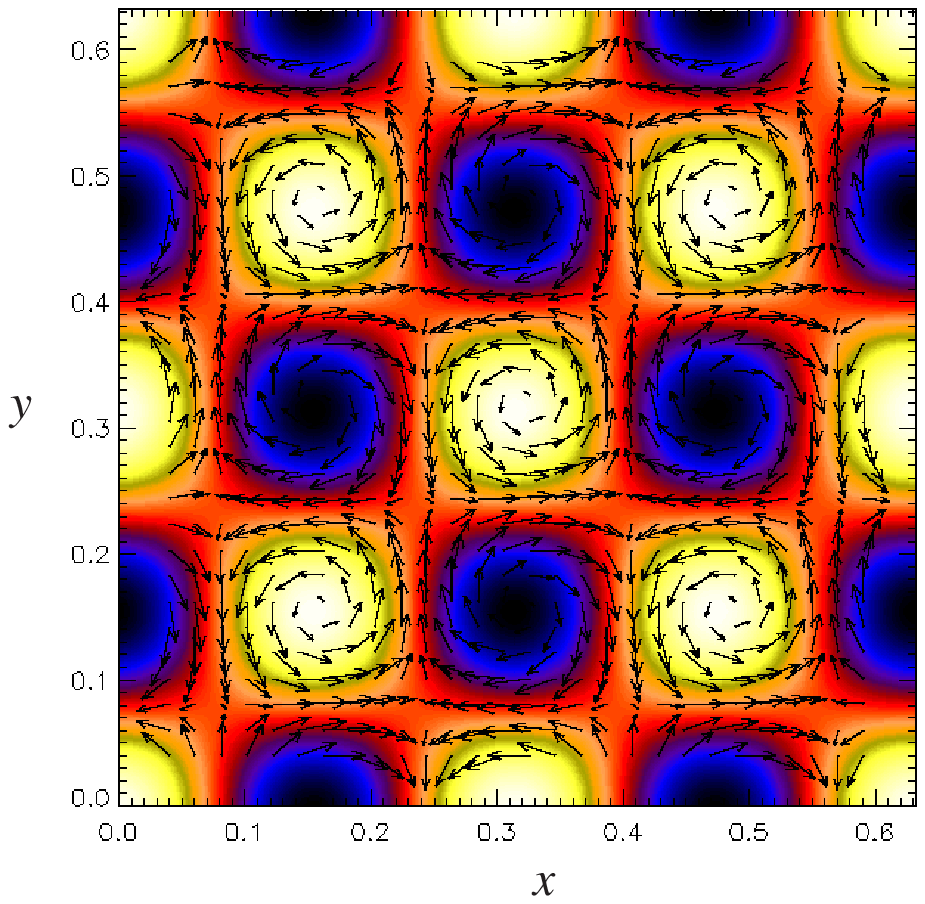} \hspace{2cm}
   \includegraphics[width=40mm]{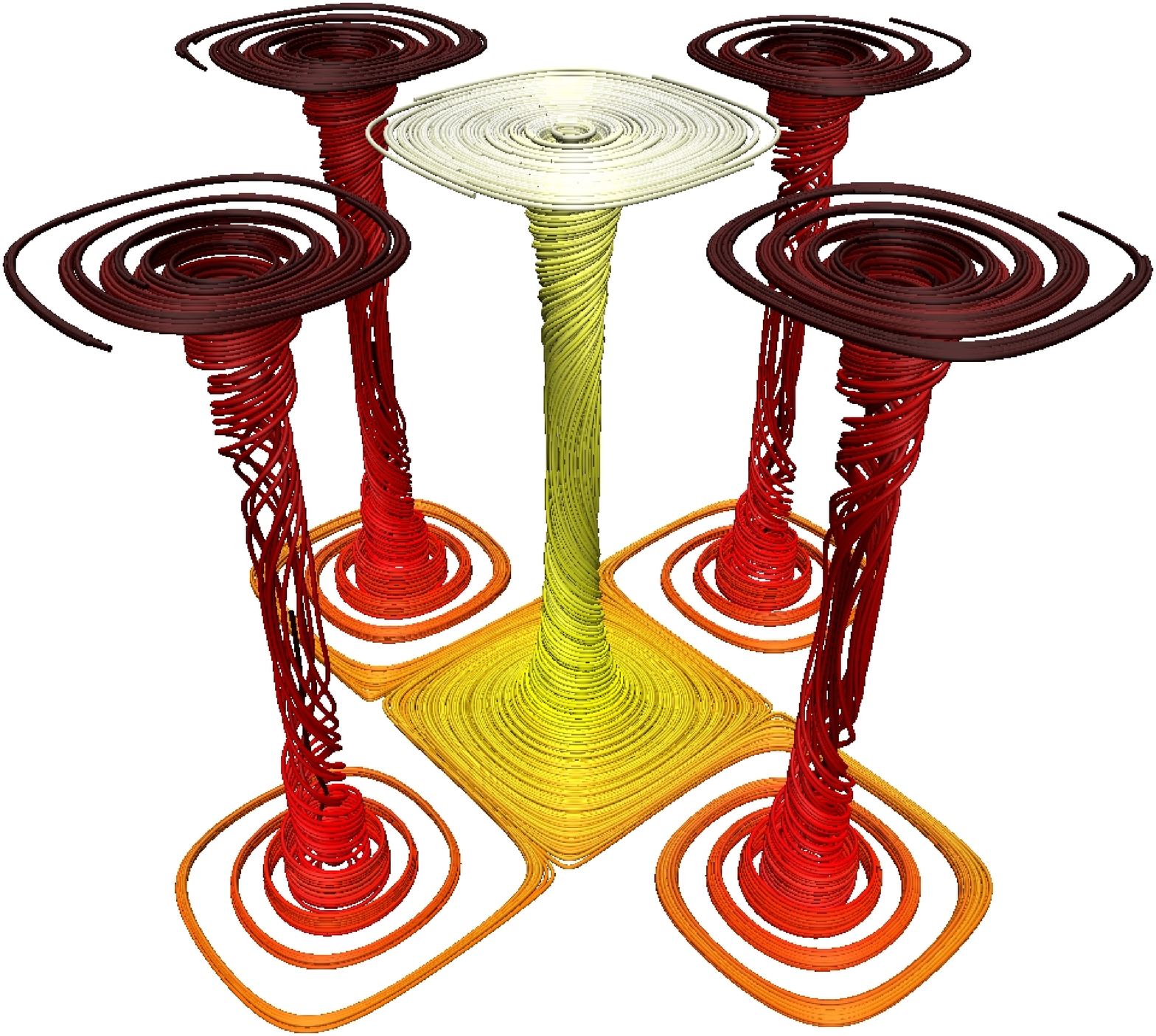} \\ \hspace{1.3cm}
   \includegraphics[width=50mm]{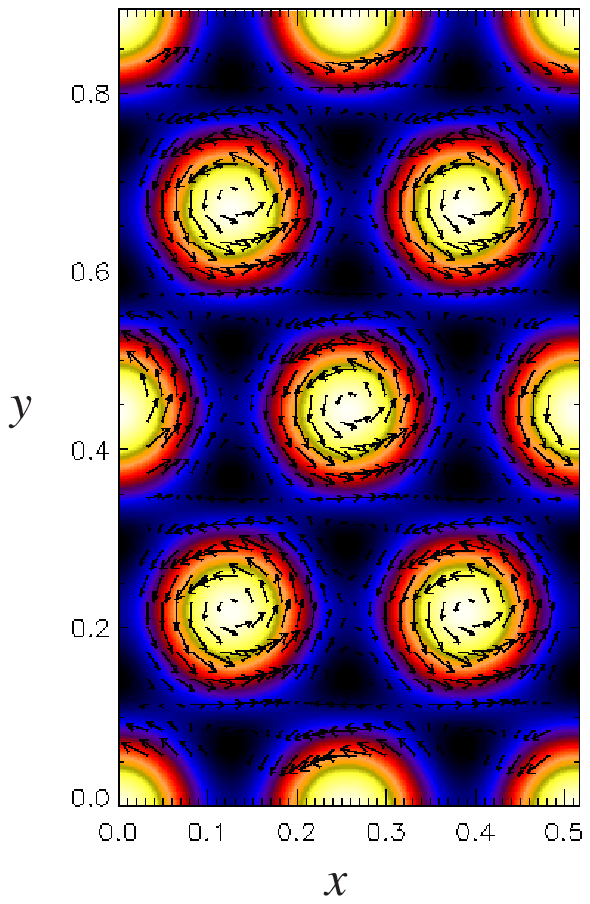}
   \includegraphics[width=60mm]{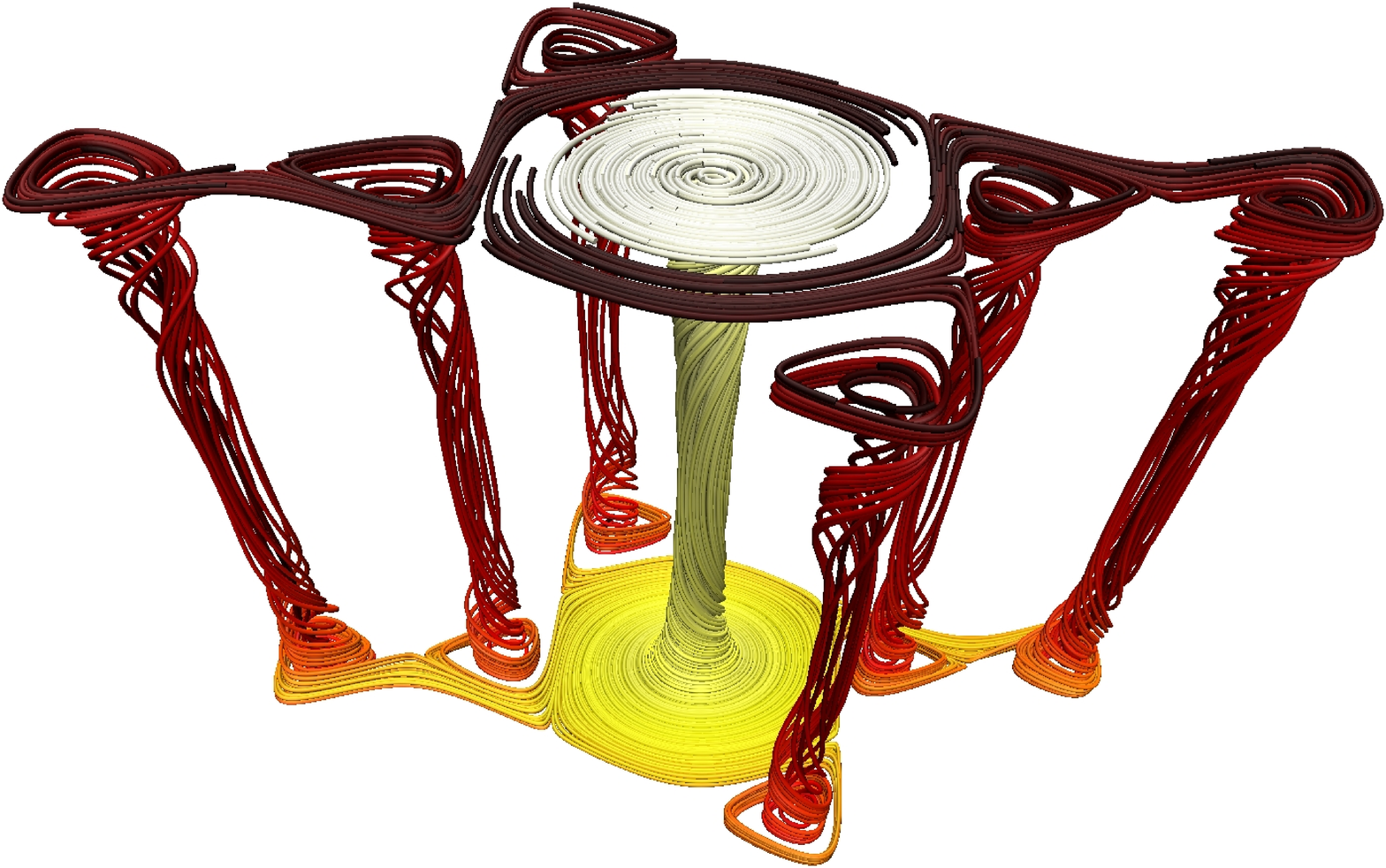}
   \caption{Left: Vertical component of the velocity in a horizontal plane located at $z=0.25$. Bright and dark colours correspond to positive and negative values respectively. The horizontal velocity field is shown with arrows. Right: Streamlines which colour depends on the time spent in the flow (the darker the longer). The starting points of the streamlines are initiated regularly in a small horizontal square grid whose size is equal to the size of a convective cell. Top: Square pattern. Bottom: Hexagonal pattern. The Taylor number is $T=10^8$ in both cases.\label{fig:ill}}
\end{figure*}

The velocity field associated with the hexagonal pattern is
\begin{multline}
\label{eq:hex1}
u_x=-\frac{\pi}{3a^2}\Big[\frac{4\pi}{\sqrt3L}\sin\frac{2\pi}{\sqrt3L}x\cos\frac{2\pi}{3L}y+\frac{4\pi}{3L}\frac{\sqrt T}{\pi^2+a^2} \\ \left(\cos\frac{2\pi}{\sqrt3L}x+2\cos\frac{2\pi}{3L}y\right)\sin\frac{2\pi}{3L}y\Big]\cos\pi z
\end{multline}
\begin{multline}
\label{eq:hex2}
u_y=-\frac{\pi}{3a^2}\Big[\frac{4\pi}{3L}\left(\cos\frac{2\pi}{\sqrt3L}x+2\cos\frac{2\pi}{3L}y\right)\sin\frac{2\pi}{3L}y- \\
\frac{4\pi}{\sqrt3L}\frac{\sqrt T}{\pi^2+a^2}\sin\frac{2\pi}{\sqrt3L}x\cos\frac{2\pi}{3L}y\Big]\cos\pi z
\end{multline}
\begin{equation}
\label{eq:hex3}
u_z=\frac13\left(2\cos\frac{2\pi}{\sqrt3L}x\cos\frac{2\pi}{3L}y+\cos\frac{4\pi}{3L}y\right)\sin\pi z \ ,
\end{equation}
where $L=4\pi/(3a)$.
The same remark applies for this velocity field.
At large Taylor numbers, we expect the second term on the right-hand side of equations \eqref{eq:hex1} and \eqref{eq:hex2} to be dominant.

\begin{figure}
   \includegraphics[width=70mm]{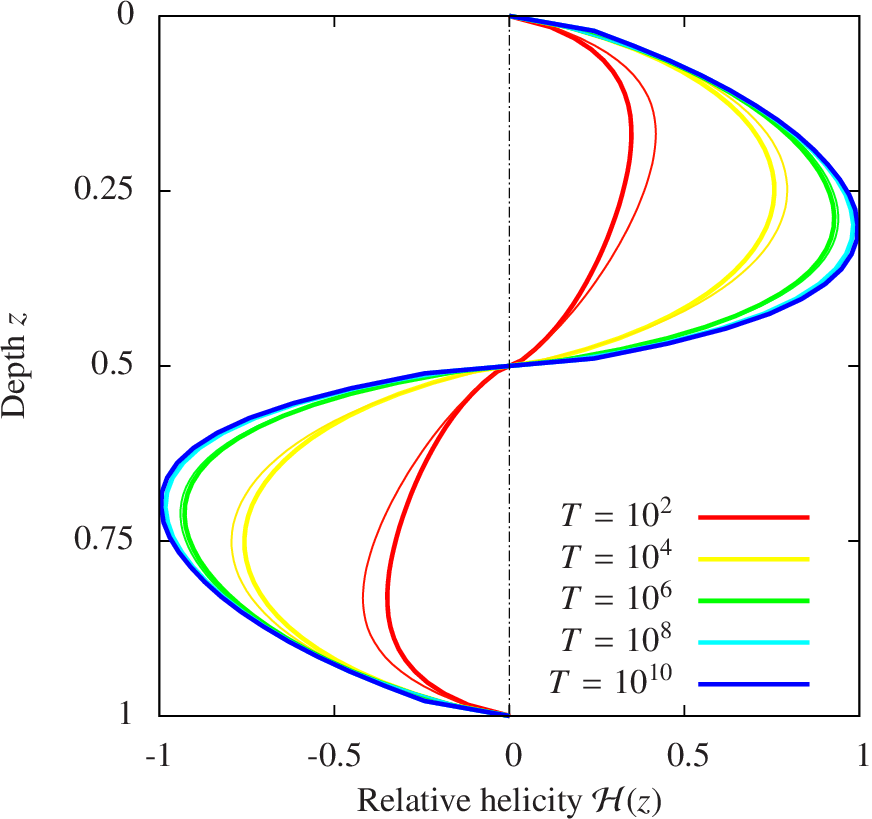}
   \caption{Horizontally averaged relative kinetic helicity as defined by equation \eqref{eq:relhel}. The results are shown for various Taylor number $T$. The thick lines correspond to the square pattern whereas the thin lines correspond to the hexagonal pattern. For $T>10^8$, both patterns converge toward the same helicity profile.\label{fig:hel}}
\end{figure}

In addition to the beautiful drawings one can find in \cite{veronis59}, which were reproduced in \cite{chan61}, we illustrate both of these velocity fields on figure \ref{fig:ill}.
Note that the flows in figure \ref{fig:ill} correspond to $T=10^8$.
The vertical component of the velocity is shown along with arrows representing the horizontal components in the horizontal plane $z=0.25$.
We also plot streamlines initiated close to the top boundary.
The symmetry between the up and down flows in the case of the square pattern is apparent, whereas a clear difference is observed in the case of the hexagonal pattern.
Note that at a particular depth and for large Taylor numbers, the square pattern flow is nearly identical to the Roberts flow \cite{roberts70,roberts72} defined by $u_x=\cos x \sin y$, $u_y=-\sin x \cos y$ and $u_z=\cos x \cos y$.
However, and contrary to the Roberts flow, the flows described by equations \eqref{eq:square1}-\eqref{eq:square3} and equations \eqref{eq:hex1}-\eqref{eq:hex3} are not maximally helical.
The relative kinetic helicity, defined by
\begin{equation}
\label{eq:relhel}
\mathcal{H}(z)=\frac{\left<\bm{u}\cdot\nabla\times\bm{u}\right>}{\left<\bm{u}^2\right>^{1/2}\left<(\nabla\times\bm{u})^2\right>^{1/2}} \ ,
\end{equation}
is presented on figure \ref{fig:hel}, where $<.>$ denotes the horizontal average over $x$ and $y$.
We plot the results for the square pattern in thick lines and for the hexagonal pattern in thin lines.
For $T>10^8$, both flows converge towards the same helicity profile.
These high Taylor number flows are nearly Beltrami (\textit{i.e.} $\nabla\times\bm{u}=\bm{u}$) for $z\approx0.28$ and $z\approx 0.72$.
Although the volume-averaged helicity is zero, these flows lack mirror-symmetry and are therefore good candidate for a mean-field type dynamo.
It is indeed known that, at infinitely large Taylor numbers, rotating convection can sustain a large-scale saturated magnetic field, both in the Boussinesq \cite{soward74} and in the anelastic \cite{mizerski2012} approximations.
Note that we focus on flows for which $T\ge10^8$ in the following.

The purpose of this paper is to study the kinematic dynamo properties of these flows by solving the induction equation
\begin{equation}
\label{eq:induction}
\frac{\partial \bm{B}}{\partial t}=\nabla\times\left(\bm{u}\times\bm{B}-\eta\nabla\times\bm{B}\right) \ ,
\end{equation}
where $\bm{u}$ is the prescribed steady velocity field given by equations \eqref{eq:square1}-\eqref{eq:square3} or equations \eqref{eq:hex1}-\eqref{eq:hex3}, and $\bm{B}$ is the magnetic field.
Both $\bm{u}$ and $\bm{B}$ are solenoidal.
We now dimensionalize lengths with the layer depth $d$, so that the dimensionless depth is unity.

In the horizontal directions, all variables are assumed to be periodic, with the same periodicity as the flow.

The upper and lower boundaries are assumed to be impermeable and stress-free, which implies that $u_{x,z}=u_{y,z}=u_z=0$ at $z=0$ (the upper boundary) and $z=1$ (the lower boundary).
We choose appropriate conditions for perfectly-conducting boundaries, which implies that $B_z = B_{x,z}=B_{y,z}=0$ at $z=0$ and $z=1$.
We also explore in section \ref{sec:bou} the effect of the magnetic boundary conditions by considering the case of a vertical field at the boundaries (the magnetic permeability of the boundaries is infinite), setting $B_x = B_y = B_{z,z} = 0$ at $z=0$ and $z=1$.

The induction equation \eqref{eq:induction} is solved using a modified version of the mixed pseudo-spectral/finite difference code that was originally described by \cite{matt95}.
Due to periodicity in the horizontal direction, horizontal derivatives are computed in Fourier space using fast Fourier transforms.
In the vertical direction, a fourth-order finite differences scheme is used, adopting an upwind stencil for the advective terms.
The time-stepping is performed by an explicit third-order Adams-Bashforth technique, with a variable time-step.
The resolution goes up to $256$ Fourier modes in each horizontal directions and $480$ grid-points in the vertical direction.
A poloidal-toroidal decomposition is used for the magnetic field in order to ensure that the field remains solenoidal.

%
%
\section{Mean-field model \label{sec:mean}}

In this section, we derive a reduced model based on mean-field theory.
The analysis performed here is closely related to the asymptotic analysis by Soward \cite{soward74} of a convectively driven magnetic dynamo in an incompressible medium, in a plane layer with strong background rotation.
Soward \cite{soward74} derived a set of nonlinear equations governing the evolution of this dynamo, and stable periodic solutions are shown to exist.
Our approach is however much simpler as we focus on the kinematic problem only.
This simplification allows us to extend the analysis to higher order than in \cite{soward74}, revealing new interesting behaviours.

It is well known that, for large Taylor numbers, the horizontal scale of the motion at the onset of the instability is of order $T^{-1/6}$. 
The parameter $\epsilon$ is therefore classically introduced \citep{soward74} and is related to the Taylor number $T$ through
\begin{equation}
\epsilon=T^{-1/6} \ .
\end{equation}

We then assume that the horizontal gradients are much larger than the vertical ones by introducing the substitution $(\partial_x,\partial_y)\rightarrow \epsilon^{-1}(\partial_x,\partial_y)\equiv\epsilon^{-1}\nabla_h$. We further assume that $\bfu=O(1)$ and that the magnetic field can be decomposed as $\bfB(z,t)+\ehalf\bfb$, where $\bfB$ is the spatial average of the magnetic field over horizontal coordinates whereas $\bfb$ is the remaining fluctuating part, which has zero horizontal average.
The time derivative scales as $\partial_t\rightarrow\ehalf\partial_t$.
If $\av{}$ denotes the horizontal average over $x$ and $y$, the mean induction equation can be written as
\begin{equation}
\partial_t \bfB=\hatzed\times\frac{\partial}{\partial z}\av{\bfu\times\bfb}+\frac{1}{\lambda}\frac{\partial^2\bfB}{\partial z^2} \ ,
\end{equation}
where $\lambda=O(1)=\epsilon^{1/2}R_m^L$ and $R_m^L=Ud/\eta$ is the large-scale magnetic Reynolds number and $U$ is a characteristic velocity.
The velocity field $\bm{u}$ will be defined later but for now, we just assume that $\nabla_h^2\bfu=-\bfu$, and that $\av{\bfu}=0$, which is verified by both square and hexagonal patterns (see equations \eqref{eq:square1}-\eqref{eq:square3} and \eqref{eq:hex1}-\eqref{eq:hex3}).
We define the small-scale magnetic Reynolds number (based on the small horizontal scale of the motion) as $R_m^S=\epsilon^{\frac12}\lambda$.
For a mean-field dynamo to operate, $R_m^L$ should be large whereas $R_m^S$ should be small, the product of these two being a constant \cite{moff78}. 
The equation for the fluctuating magnetic field is
\begin{multline}
\label{eq:littleb}
\epsilon\frac{\partial\bm{b}}{\partial t}=\epsilon^{-1}\bfB\cdot\nabla_h\bfu- u_z\frac{\partial\bm{B}}{\partial z} \\ +\left(\epsilon^{-\frac{1}{2}}\nabla_h+\ehalf\hatzed\frac{\partial}{\partial z}\right)\times\left(\bfu\times\bfb-\av{\bfu\times\bfb}\right) \\ +\epsilon^{-1}\frac{1}{\lambda}\left(\nabla_h^2+\epsilon^2\frac{\partial^2}{\partial z^2}\right)\bfb \ .
\end{multline}
We now expand the fluctuating magnetic field as $\bfb=\bfb_0+\ehalf\bfb_1$.
At leading order, equation \eqref{eq:littleb} gives
\begin{equation}
0=\bfB\cdot\nabla_h\bfu+\frac{1}{\lambda} \nabla_h^2\bfb_0 \ .
\end{equation}
Hence $\bfb_0=\lambda\bfB\cdot\nabla_h\bfu$.
The mean electromotive force is, at first order,
\begin{equation}
\bm{\calE}_0\equiv\av{\bfu\times\bfb_0}=\lambda\left<\bfu\times(\bfB\cdot\nabla_h)\bfu\right> \ .
\end{equation}
At the next order, equation \eqref{eq:littleb} gives
\begin{equation}
0=\nabla_h\times(\bfu\times\bfb_0-\av{\bfu\times\bfb_0})+\frac{1}{\lambda}\nabla_h^2\bfb_1 \ ,
\end{equation}
and the correction to the mean electromotive force is
\begin{equation}
\begin{split}
\bm{\calE}_1\equiv\av{\bfu\times\bfb_1}&=-\av{\nabla_h^2\bfu\times\bfb_1}=-\av{\bfu\times\nabla_h^2\bfb_1}\\
&=\lambda^2\av{\bfu\times\nabla_h\times(\bfu\times(\bfB\cdot\nabla_h)\bfu)} \ .
\end{split}
\end{equation}
To simplify this expression, consider two vector fields $\bfP$ and $\bfQ$. Then,
\begin{equation}
\label{eq:vect}
\av{\bfP\times\nabla_h\times\bfQ}_x=\av{Q_x\nabla_h\cdot\bfP-Q_j\left(\frac{\partial P_j}{\partial x}\right)} \ .
\end{equation}
\begin{figure*}
      \includegraphics[width=65mm]{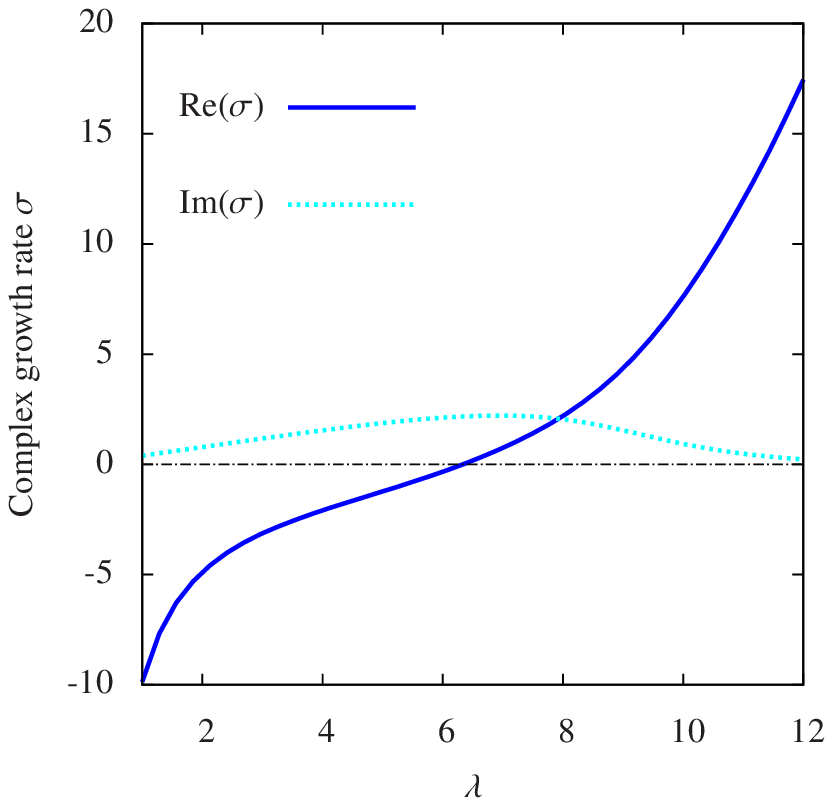} \quad \quad
      \includegraphics[width=82mm]{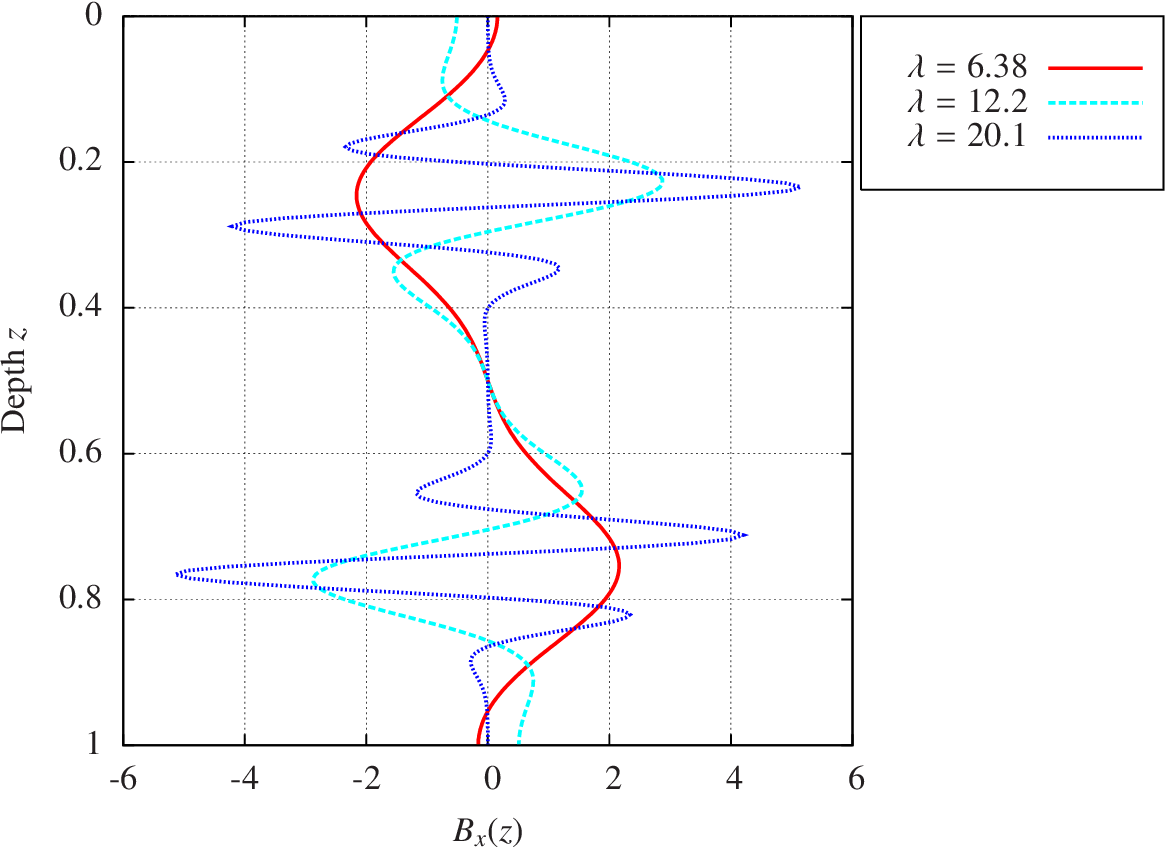}
    \caption{Results from the mean-field equation \eqref{eq:evps} corresponding to the square pattern. Left: real and imaginary parts of the growth rate $\sigma$ versus $\lambda$. Right: Eigenfunctions $B_x(z)$ for $\lambda=6.38$ (just above the onset of dynamo action), $\lambda=12.2$ and $\lambda=20.1$. The eigenfunctions have been normalized so that $\int_0^1|B_x(z)|\textrm{d}z=1$.}
    \label{fig:evps}
\end{figure*}
By taking $\bfP=\bfu$, the first term in the right-hand side of equation \eqref{eq:vect} is $O(\epsilon)$ and can therefore be neglected.
Thus the $x$-component of $\bm{\calE}_1$ is 
\begin{equation}
-\lambda^2\left\langle(\bfu\times(\bfB\cdot\nabla_h)\bfu)\cdot\frac{\partial \bfu}{\partial x}\right\rangle=-\lambda^2B_y\left\langle\bfu\times\frac{\partial\bfu}{\partial y}\cdot\frac{\partial \bfu}{\partial x}\right\rangle \ ,
\end{equation}
while the $y$-component is
\begin{equation}
+\lambda^2B_x\left\langle\bfu\times\frac{\partial\bfu}{\partial y}\cdot\frac{\partial \bfu}{\partial x}\right\rangle \ .
\end{equation}
The $z$-component is irrelevant as it will disappear when the curl is taken in the mean equation, so we assume $\bm{\calE}_1$ is horizontal.
This is a result of the fast rotation considered here \citep{ruediger1978}.
Then clearly $\bm{\calE}_1$ is perpendicular to $\bfB$ so that $\bm{\calE}_1=\bfV\times\bfB$, where 
\begin{align}
\bfV & =\lambda^2\hatzed\left\langle\bfu\times\frac{\partial\bfu}{\partial y}\cdot\frac{\partial \bfu}{\partial x}\right\rangle \\ & \label{eq:pumping} =3\lambda^2\hatzed\left\langle u_z\left(\frac{\partial u_x}{\partial y}\frac{\partial u_y}{\partial x}-\frac{\partial u_x}{\partial x}\frac{\partial u_y}{\partial y}\right)\right\rangle \ .
\end{align}
$\bfV$ is a pumping velocity, and corresponds to the off-diagonal terms of the classical $\alpha$ tensor of mean-field electrodynamics.

We now specialise to velocity fields that mimic that found by Veronis \cite{veronis59}.
Let us write the velocity field as
\begin{equation}
\bfu=(\nabla_h \phi\times\hatzed)\cos\pi z+\phi\hatzed \sin\pi z \ .
\end{equation}
In the case of the square pattern, we can choose
\begin{equation}
\phi=\cos x \cos y \ .
\end{equation}
Note that the resulting flow is not solenoidal, but the correction necessary to recover equations \eqref{eq:square1}-\eqref{eq:square3} is $O(\epsilon)$ and so can be neglected here.
In that case, the pumping velocity $\bfV$ is zero and the mean electromotive force reduces to its first order term
\begin{equation}
\label{eq:emfs}
\bm{\calE}_0=-\frac{1}{2}\lambda\bfB\sin \pi z\cos \pi z \ ,
\end{equation}
so that the equation to solve is
\begin{equation}
\label{eq:evps}
\frac{\partial\bfB}{\partial t}=-\frac{1}{2}\lambda\hatzed\times\frac{\partial}{\partial z}\Big[\bfB\sin(\pi z)\cos(\pi z)\Big]+\frac{1}{\lambda}\frac{\partial^2\bfB}{\partial z^2} \ .
\end{equation}
Note that the equation does not depend on $\epsilon$, but we require $\epsilon\ll1$.
\begin{figure*}
      \includegraphics[width=63mm]{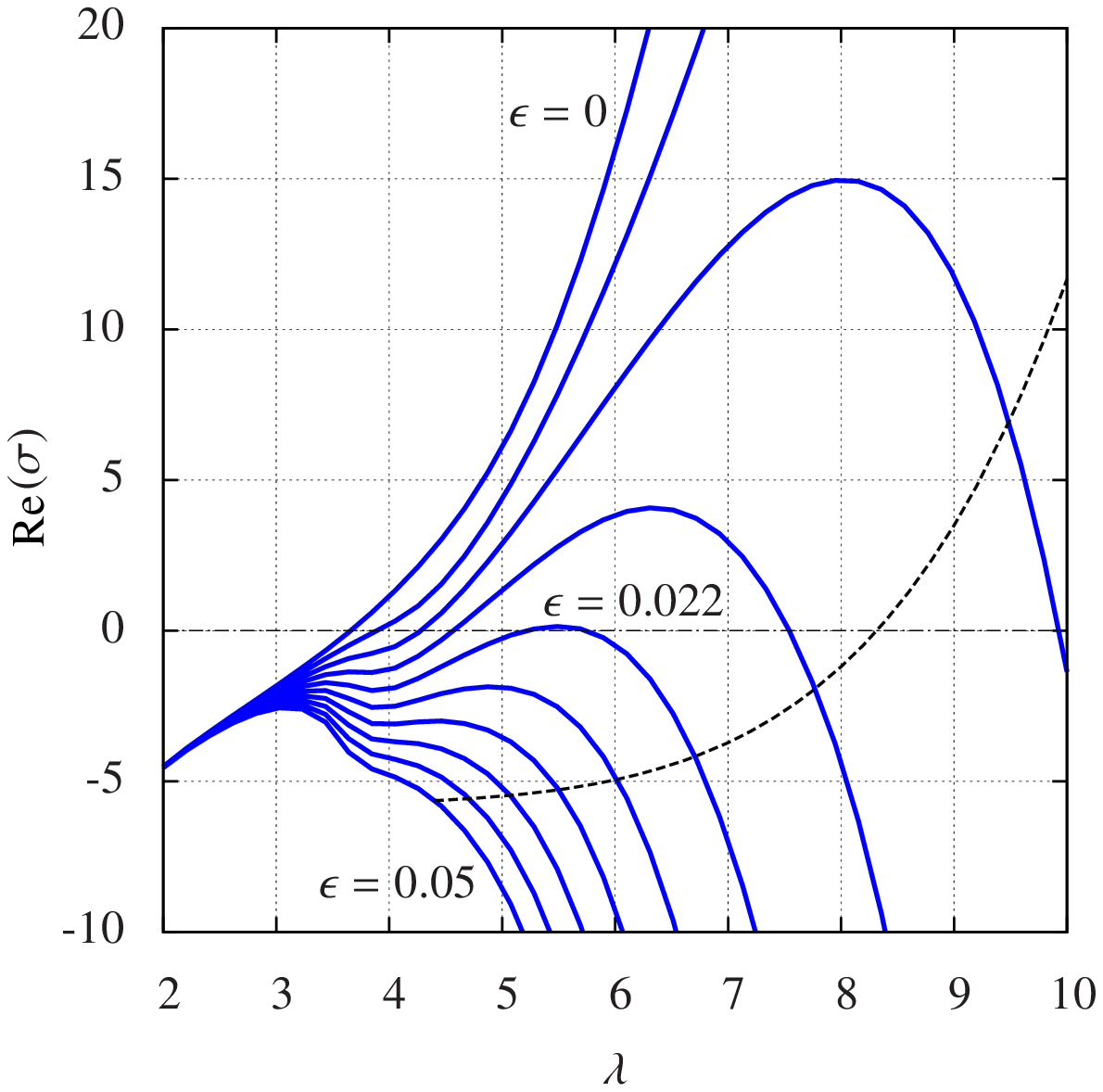} \quad \quad
      \includegraphics[width=86mm]{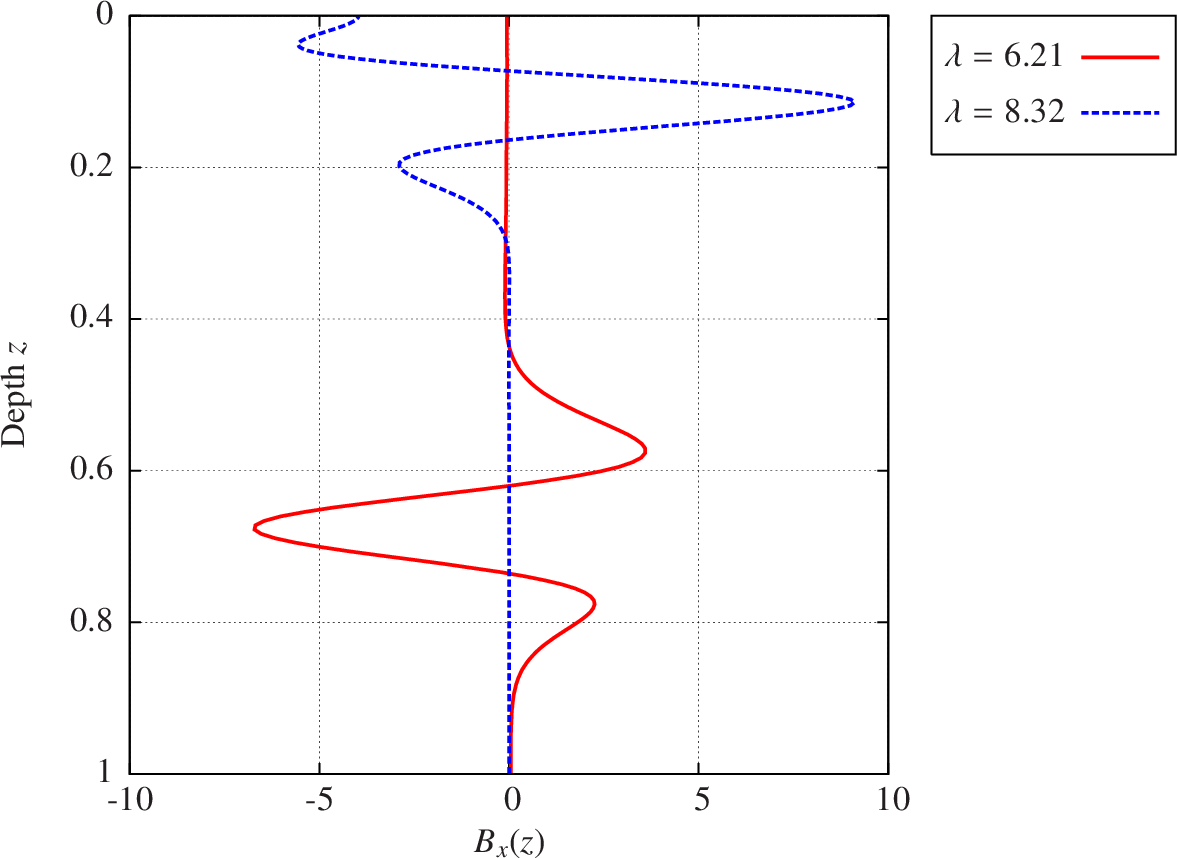}
    \caption{Results from the mean-field equation \eqref{eq:evph} corresponding to the hexagonal pattern. Left: real part of the growth rate $\sigma$ versus $\lambda$ for different $\epsilon$. The dashed curve corresponds to the location of $R_m^S=\epsilon^{1/2}\lambda=1$. Points above this line are beyond the domain of validity of mean-field theory. Right: Eigenfunctions $B_x(z)$ for $\epsilon=0.0192$ ($T=2\times10^{10}$) and two different values of $\lambda$. The eigenfunctions have been normalized so that $\int_0^1|B_x(z)|\textrm{d}z=1$. The case $\lambda=6.21$ is a dynamo ($\textrm{Re}(\sigma)=1.67$) whereas the case $\lambda=8.32$ is not ($\textrm{Re}(\sigma)=-19.39$).}
    \label{fig:evph}
\end{figure*}

In the case of the hexagonal pattern, we can choose
\begin{multline}
\label{eq:veronis}
\phi(x,y)=\cos x+\cos\left(-\half x+\textstyle{\frac{\sqrt{3}}{2}}y\right) \\ +\cos\left(-\half x-\textstyle{\frac{\sqrt{3}}{2}}y\right) \ .
\end{multline}
Again, this flow is not solenoidal, but the correction necessary to recover equations \eqref{eq:hex1}-\eqref{eq:hex3} is also $O(\epsilon)$.
In that case, the pumping velocity does not vanish, and equation \eqref{eq:pumping} can be rewritten after some algebra as 
\begin{align}
\bfV = & 3\lambda^2\hatzed\sin(\pi z)\cos^2(\pi z)\left\langle\phi\left[\left(\frac{\partial^2 \phi}{\partial x\partial y}\right)^2-\frac{\partial^2\phi}{\partial x^2}\frac{\partial^2\phi}{\partial y^2}\right]\right\rangle \\ \label{eq:pumping2} = & -\frac{27}{16}\lambda^2\hatzed\sin(\pi z)\cos^2(\pi z).
\end{align}
A similar calculation for the mean electromotive force at first order gives
\begin{equation}
\bm{\calE}_0=-\frac{3}{2}\lambda\bfB\sin\pi z\cos\pi z \ .
\end{equation}
Finally, the equation to solve is
\begin{multline}
\label{eq:evph}
\frac{\partial\bfB}{\partial t}=-\frac{3}{2}\lambda\hatzed\times\frac{\partial}{\partial z}\Big[\bfB\sin(\pi z)\cos(\pi z)\Big] \\ +\frac{27}{16}\lambda^2\epsilon^{1/2}\frac{\partial}{\partial z}\Big[\bfB\sin(\pi z)\cos^2(\pi z)\Big]+\frac{1}{\lambda}\frac{\partial^2\bfB}{\partial z^2} \ .
\end{multline}
Note that equation \eqref{eq:evph} involves a term depending on our small parameter $\epsilon$.
This is the only additional term at that order, and such a term is exactly zero in the case of the square pattern (see equation \eqref{eq:evps}).
Higher order terms are neglected as they do not provide further insights into the problem, and are similar for both flows.

We now look for a solution of equations \eqref{eq:evps} and \eqref{eq:evph}. The horizontally-averaged magnetic field $(B_x(z,t),B_y(z,t))$ is written in the form $\bm{B}(z)e^{\sigma t}$, where $\sigma$ is the complex growth rate.
The functions $B_x$ and $B_y$ are represented by their discretized values at the Gauss-Lobatto collocation nodes
\begin{equation}
z_i=\cos\left(\frac{i\pi}{N}\right) \ , \quad 0 \le i \le N
\end{equation}
where $N$ is the Chebyshev truncation order.
Each differential equation is then represented at each of the collocation nodes, using the first and second Chebyshev collocation derivative matrices.
The boundary conditions are represented at the two boundary points $z=0$ and $z=1$, again using the Chebyshev collocation derivative where needed.
The following results are derived using a Chebyshev truncation order of $N=256$.
The generalized eigenvalue problem associated with equations \eqref{eq:evps} and \eqref{eq:evph} can be written as
\begin{equation}
M \bm{B}=\sigma P \bm{B} \ ,
\end{equation}
where $M$ is the matrix associated with the discretized linear operators, $P$ is the matrix associated with the boundary conditions and $\bm{B}$ is a vector containing the variables $B_x$ and $B_y$ at the collocation points.
We recall here that the boundary conditions correspond to a perfectly conducting medium at $z=0$ and $z=1$, which imposes $B_{x,z}=B_{y,z}=0$.

This generalized eigenvalue problem is solved for $M$ and $P$ using the following method.
First, the matrix $P$ is transformed in an upper-diagonal matrix.
We then reduce the pair of real matrices $(M,P)$ to a generalized upper Hessenberg form using orthogonal transformations.
The eigenvalues are finally computed using the double-shift QZ method.
These different steps are performed using the relevant routines from the Linear Algebra PACKage library.
We then select the eigenfunction associated with the largest real eigenvalue $\textrm{Re}(\sigma)$, with the additional constraint that the total horizontal magnetic fluxes are zero:
\begin{equation}
\label{eq:consflux}
\int_0^1 B_x(z) \textrm{d}z=0 \quad \textrm{and} \quad \int_0^1 B_y(z) \textrm{d}z=0 \ .
\end{equation}
This constraint must be respected at all times due to the combination of horizontal periodicity and perfectly conducting boundary conditions at $z=0$ and $z=1$.
\begin{figure*}
      \includegraphics[width=58mm]{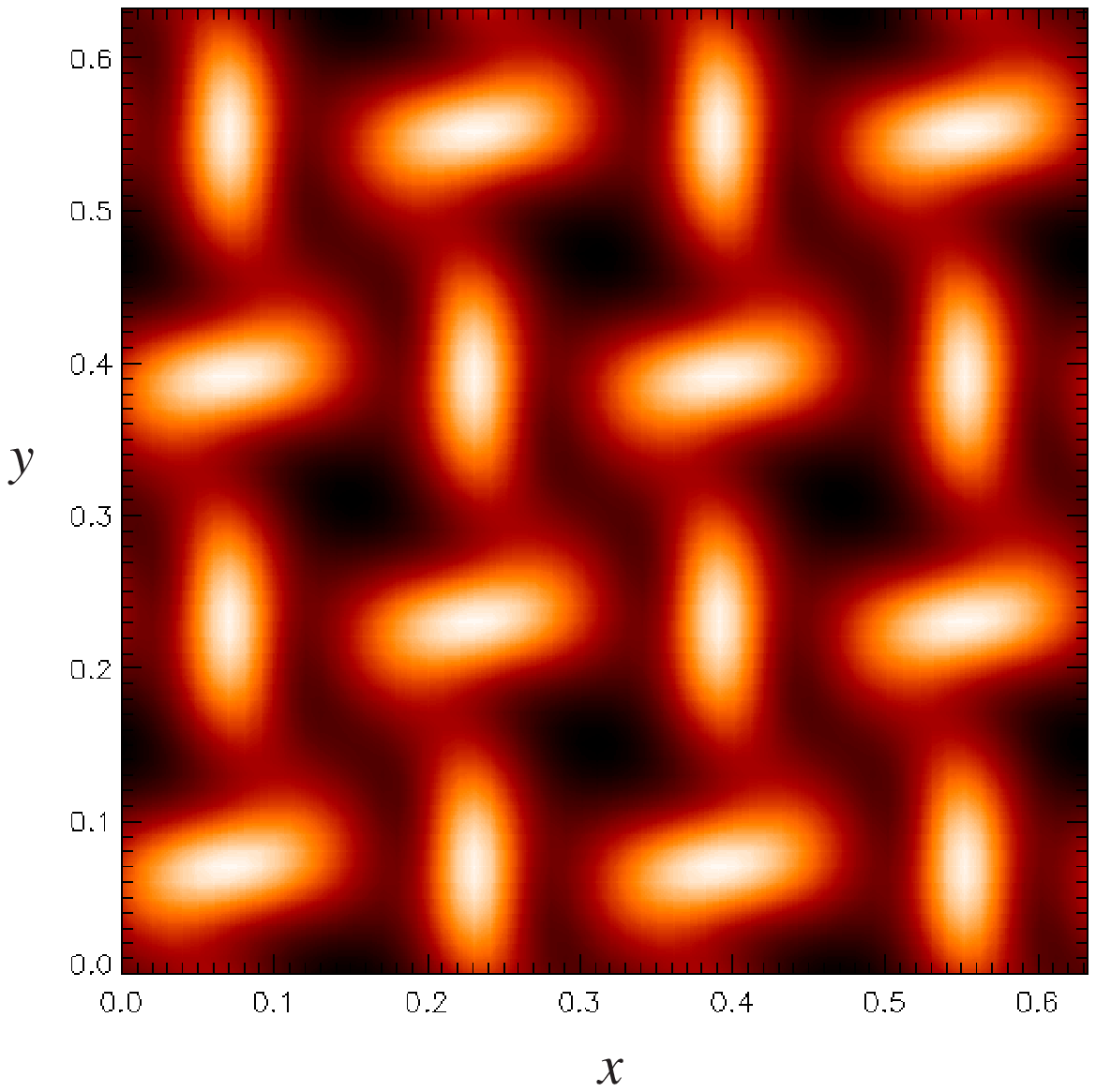}
      \includegraphics[width=58mm]{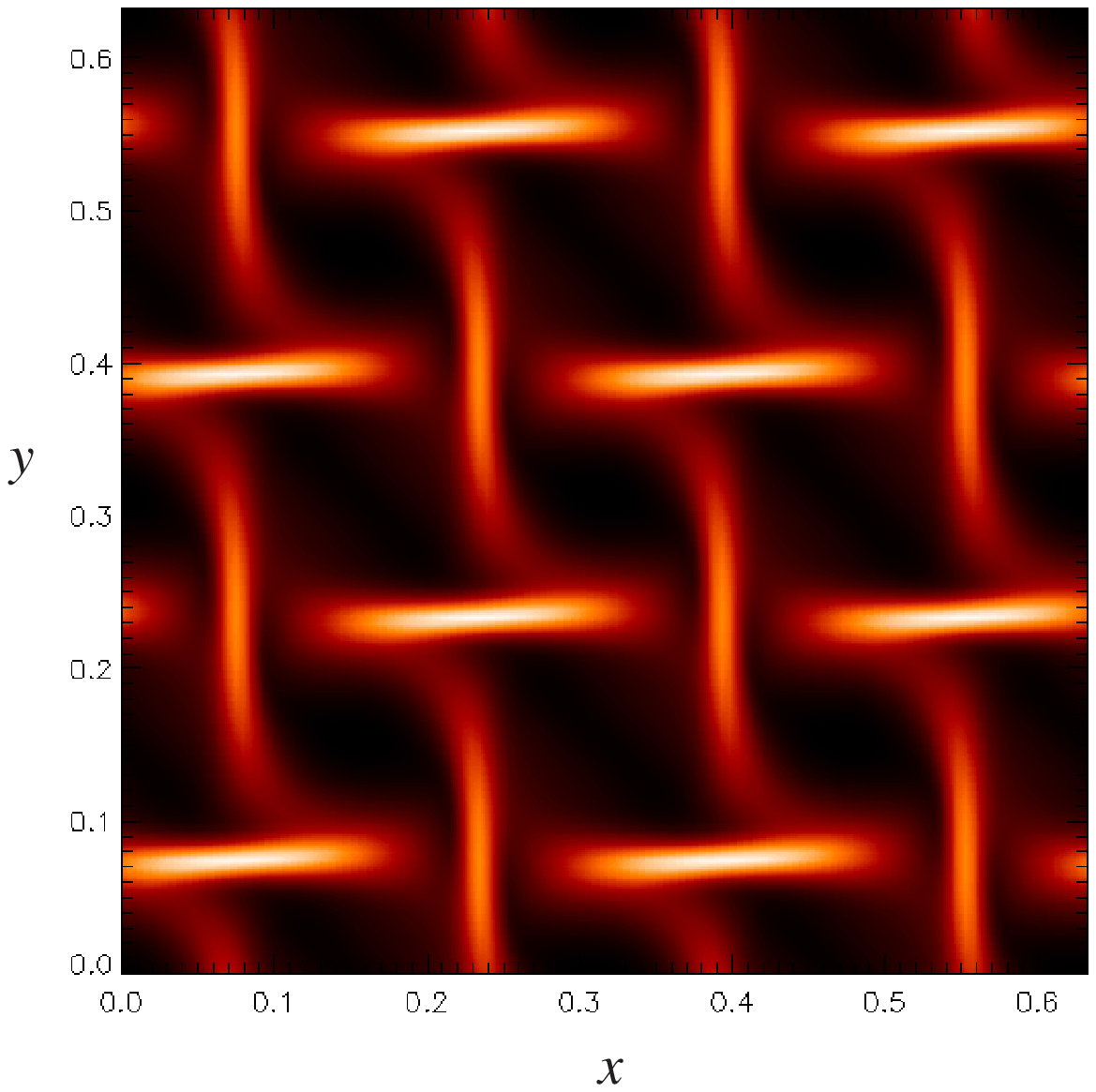}
      \includegraphics[width=58mm]{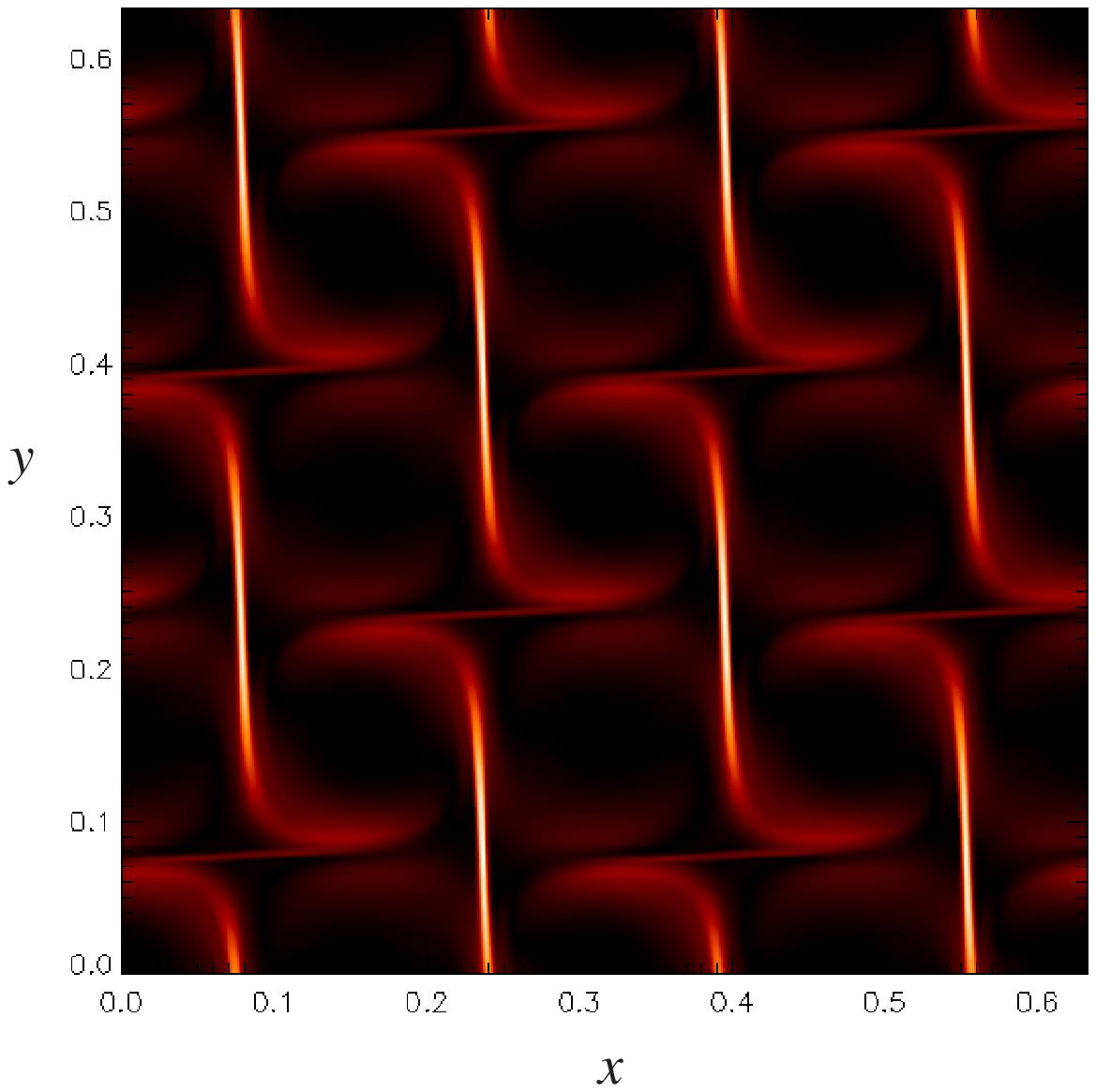}
    \caption{Magnetic energy $\bm{B}^2$ in a horizontal plane located at $z=0.25$. For the sake of clarity, the aspect ratio has been multiplied by two by plotting four copies of the domain side by side. Bright and dark tones correspond to opposite polarity. From left to right, $\eta=10^{-2}$ (close to the critical value for dynamo action), $\eta=10^{-3}$ and $\eta=5\times10^{-5}$ (small-scale dynamo action is possible). The horizontally-averaged magnetic field is oriented in the same direction in all cases.}
    \label{fig:bx}
\end{figure*}

Let us first discuss the results associated with the square pattern and equation \eqref{eq:evps}.
The only relevant parameter is here $\lambda$, which is related to the large-scale magnetic Reynolds number by $R_m^L=\epsilon^{-1/2}\lambda$.
In the following, we vary $\lambda$ between $1$ and $12$.
The real and imaginary parts of the eigenvalues are plotted on figure \ref{fig:evps}.
The critical value of $\lambda$ for dynamo action appears to be $\lambda_{\mathrm{crit}}\approx6.335$.
Using the same scaling as in \cite{soward74}, this corresponds to $\Lambda_{\mathrm{crit}}=\lambda_{\mathrm{crit}}^2/(8\pi)\approx1.597$, which is consistent with the value quoted in the same paper (see p.623 of \cite{soward74}).
The imaginary part of the growth rate is always positive with a maximum at $\lambda=6.93$, slightly after the onset for dynamo, and decays to zero at large $\lambda$.
We therefore expect the dynamo to be oscillatory at onset.
The eigenfunctions for $B_x$ are also shown on figure \ref{fig:evps} for various values of $\lambda$.
Note that dynamo action is confined in the upper and lower halves of the domain, where the relative helicity and the mean electromotive force are extremal.
The eigenfunctions are anti-symmetric with respect to the mid-layer, as it is the case for the relative helicity (see figure \ref{fig:hel}) and the mean electromotive force (see equation \eqref{eq:emfs}).
As $\lambda$ is increasing, the small-scale magnetic Reynolds number $R_m^S=\epsilon^{\frac12}\lambda$ increases until mean-field theory is not applicable anymore.
As the Taylor number increases for a fixed $\lambda$, $\epsilon$ and $R_m^S$ decrease so that the range of applicability of mean-field theory increases.
From this model, we can derive a minimal Taylor number for large-scale dynamo to occur.
By assuming that $R_m^S=\epsilon^{\frac12}\lambda=1$ and using the critical value of $\lambda$, one finds a limit Taylor number of $T\approx4\times10^{9}$.
Below this value, mean-field theory is not applicable at the onset of the mean-field dynamo, so that we cannot conclude as to its existence.

We now discuss the results associated with the hexagonal pattern and equation \eqref{eq:evph}.
We vary the parameter $\epsilon$ between $0$ and $5\times10^{-2}$, which corresponds to $T=\infty$ and $T=6.4\times10^7$ respectively.
Note that in the case $\epsilon=0$, the pumping velocity is zero, and equations \eqref{eq:evps} and \eqref{eq:evph} are identical apart from the numerical coefficient in front of the mean electromotive force.
The parameter $\lambda$ is varied between $1$ and $10$.
\begin{table}
 \begin{center}
\def~{\hphantom{0}}
 \begin{tabular}{cccccc}
   \hline
   $T$ & $\epsilon$ & $a$ & $\lambda_x^S=\lambda_y^S$ & $\lambda_x^H$ & $\lambda_y^H$ \\
   \hline
   $10^8$ \hspace{1mm}&\hspace{1mm} $0.0464$ \hspace{1mm}&\hspace{1mm} $28.11$ \hspace{1mm}&\hspace{1mm} $0.316$ \hspace{1mm}&\hspace{1mm} $0.258$ \hspace{1mm}&\hspace{1mm} $0.447$ \\
   $10^{10}$ \hspace{1mm}&\hspace{1mm} $0.0215$ \hspace{1mm}&\hspace{1mm} $60.56$ \hspace{1mm}&\hspace{1mm} $0.147$ \hspace{1mm}&\hspace{1mm} $0.12$ \hspace{1mm}&\hspace{1mm} $0.207$ \\
   $2\times10^{10}$ \hspace{1mm}&\hspace{1mm} $0.0192$ \hspace{1mm}&\hspace{1mm} $67.98$ \hspace{1mm}&\hspace{1mm} $0.131$ \hspace{1mm}&\hspace{1mm} $0.107$ \hspace{1mm}&\hspace{1mm} $0.185$ \\
   $10^{12}$ \hspace{1mm}&\hspace{1mm} $0.01$ \hspace{1mm}&\hspace{1mm} $130.48$ \hspace{1mm}&\hspace{1mm} $0.068$ \hspace{1mm}&\hspace{1mm} $0.056$ \hspace{1mm}&\hspace{1mm} $0.096$ \\
   \hline
  \end{tabular}
\caption{Summary of the parameter values for different Taylor number $T$. $\epsilon$ is equal to $T^{-1/6}$. $a$ is the most unstable horizontal wave number at onset. $\lambda_x$ and $\lambda_y$ are the aspect ratios of the numerical domain. The superscripts $S$ and $H$ correspond to square and hexagonal patterns respectively. In the case of the hexagonal pattern, we have $\lambda_y^H=\sqrt3\lambda_x^H$.\label{tab:one}} 
 \end{center}
\end{table}

We show on figure \ref{fig:evph} the evolution of $\textrm{Re}(\sigma)$ with $\lambda$ for different $\epsilon$.
When $\epsilon=0$, the only remaining term in equation \eqref{eq:evph} is the $\alpha$-effect so that a mean-field dynamo is expected at sufficiently large $\lambda$ as in the square pattern case.
This is indeed observed and dynamo action is observed for $\lambda>3.66$.
As $\epsilon$ increases, the real part of the growth rate $\textrm{Re}(\sigma)$ decreases, up to the point where no mean-field dynamo is observed for approximately $\epsilon\approx0.025$, which corresponds to a critical Taylor number of approximately $T\approx4\times10^{9}$.
Below this critical value, no large-scale dynamo is possible.
Note that this reduction in the growth rate only happens at sufficiently large values of $\lambda$.
The imaginary part of the growth rate is monotonously increasing as $\lambda$ is increasing (not shown).
On figure \ref{fig:evph}, we also show as a dashed curve the location where the small-scale magnetic Reynolds number $R_m^S=\epsilon^{1/2}\lambda$ is equal to unity, which defines the upper limit of validity of mean-field theory.
The reduction in the growth rate, and the eventual disappearance of any mean-field dynamo action for sufficiently large $\epsilon$, is well within the range of validity of the model.
For $R_m^S>1$, small-scale dynamo action might be possible, and the current mean-field approach is irrelevant.
The reason why dynamo action is less efficient as $\epsilon$ increases can be understood by looking at the eigenfunctions for the horizontally-averaged magnetic field.
We show in figure \ref{fig:evph} the eigenfunctions associated with $B_x$ for a fixed value of $\epsilon=0.0192$ which corresponds to $T=2\times10^{10}$. The results are shown for $\lambda=6.21$ and $\lambda=8.32$.
In the first case, the pumping velocity is weak so that the resulting eigenfunction is not symmetric with respect to $z=0.5$ but there is still a strong mean-field in one of the two regions of high helicity.
Dynamo action is in that case possible (see the left part of figure \ref{fig:evph}).
For $\lambda=8.32$ however, the pumping velocity is dominant and the magnetic flux is advected vertically, away from region of efficient $\alpha$-effect, and dynamo action disappears.
Note that the pumping velocity is directed upward along $-\hat{\bm{z}}$, see equation \eqref{eq:pumping2}.
While the flow is more complicated in our case due to the presence of rotation, our results are qualitatively similar to the ones presented by \cite{droby74} (see their figure 4 for example).
Of course, the mean-field model is only valid when the small-scale magnetic Reynolds number $\epsilon^{1/2}\lambda$ is small compared to unity, so that small-scale dynamo might still be possible at larger values of $Rm$. It is however not possible to address this aspect of the problem with the current model. 
Note also that this simple model does not take into account turbulent diffusion terms or other $O(\epsilon)$ corrections, but as will become apparent in the following, it is enough to capture the essential properties of the kinematic dynamo action in such flows.
%
%
\section{Square pattern\label{sec:square}}
\begin{figure}
      \hspace{4mm}\includegraphics[width=75mm]{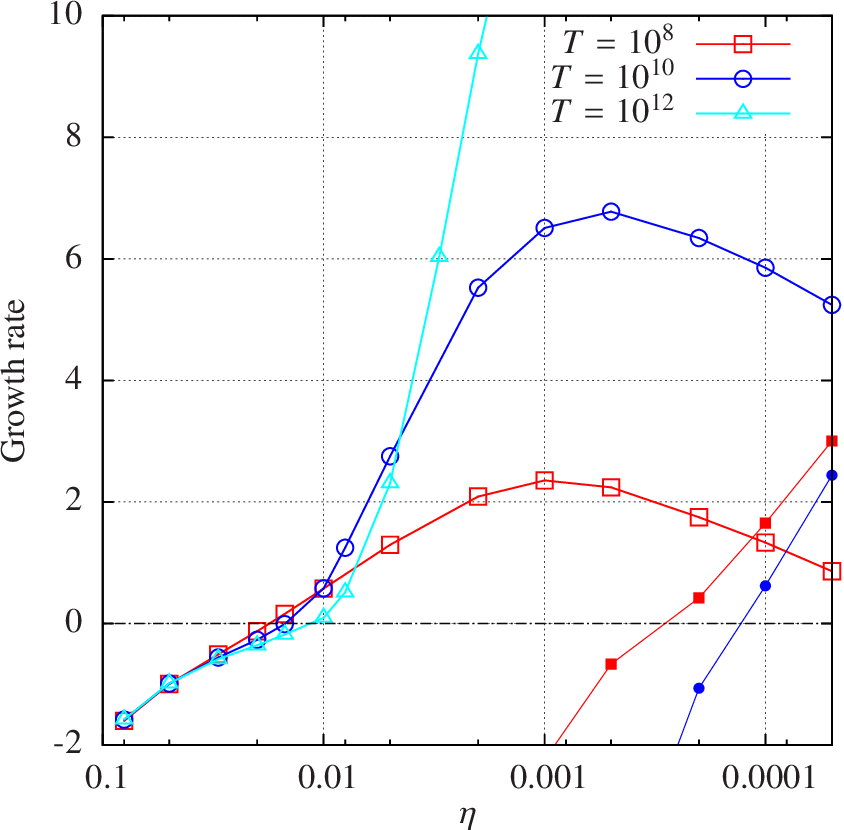}\\
      \includegraphics[width=80mm]{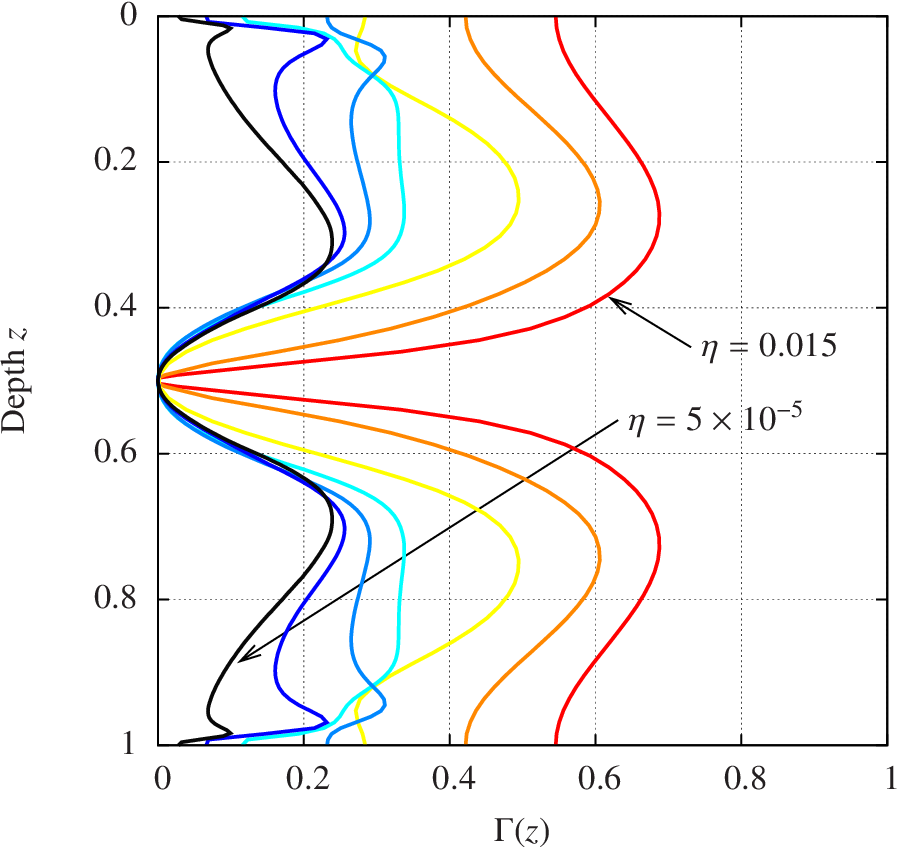}
    \caption{Top: Growth rate of the magnetic energy versus $\lambda$. The large empty symbols correspond to the full simulations whereas the small full symbols correspond to cases where the mean induction equation \eqref{eq:meanfield} is neglected. Bottom: $\Gamma(z)$, as defined by equation \eqref{eq:gamma}, for the square pattern at $T=10^8$, and for different values of $\eta$.}
    \label{fig:kinmag}
\end{figure}
In addition to confirming the predictions of the reduced mean-field model, the purpose of this section is to explore the large magnetic Reynolds number regime, for which mean-field theory is not applicable.
The induction equation \eqref{eq:induction} is now fully solved in three dimensions using the numerical scheme described in section \ref{sec:model}.
We focus in this section on the flow defined by equations \eqref{eq:square1}-\eqref{eq:square3} and corresponding to a square pattern.
From the mean-field model described in the previous section, we expect a large-scale dynamo at onset.
We consider three different Taylor numbers: $T=10^8$, $T=10^{10}$ and $T=10^{12}$.
These flows are all characterized by $\epsilon\ll1$, but we nevertheless keep all the terms in equations \eqref{eq:square1}-\eqref{eq:square3} so that the flow is rigorously incompressible.
This will lead to quantitative differences with the previously studied mean-field model for which $O(\epsilon)$ terms were neglected.
We consider the case $T=10^8$ as it is smaller than the critical value of $T\approx4\times10^9$ predicted by the mean-field model for the existence of a self-consistent large-scale dynamo. It should therefore allow us to study the behaviour of the dynamo as mean-field theory becomes gradually less and less applicable.
The cases $T=10^{10}$ and $T=10^{12}$ should be consistent with the mean-field model on a wider range of parameters.
Since the flow is periodic, we restrict our numerical solution to have the same periodicity by fixing the aspect ratio to be $\lambda_x=\lambda_y=2\sqrt2\pi/a$.
We considered numerical simulations with larger aspect ratios in order to allow for spatially-modulated solutions, but we didn't find any.
The corresponding aspect ratios can be found on Table~\ref{tab:one}.
The magnetic field is initialised with a small perturbation with zero net horizontal flux.
We vary the magnetic diffusivity from $\eta=10^{-1}$ down to $\eta=5\times10^{-5}$.
As the diffusivity is reduced, we also increase the numerical resolution.
For the case $\eta=10^{-1}$, a resolution of $32^2\times48$ is sufficient whereas the case $\eta=5\times10^{-5}$ requires a resolution of $256^2\times 480$.
\begin{figure}
      \includegraphics[width=90mm]{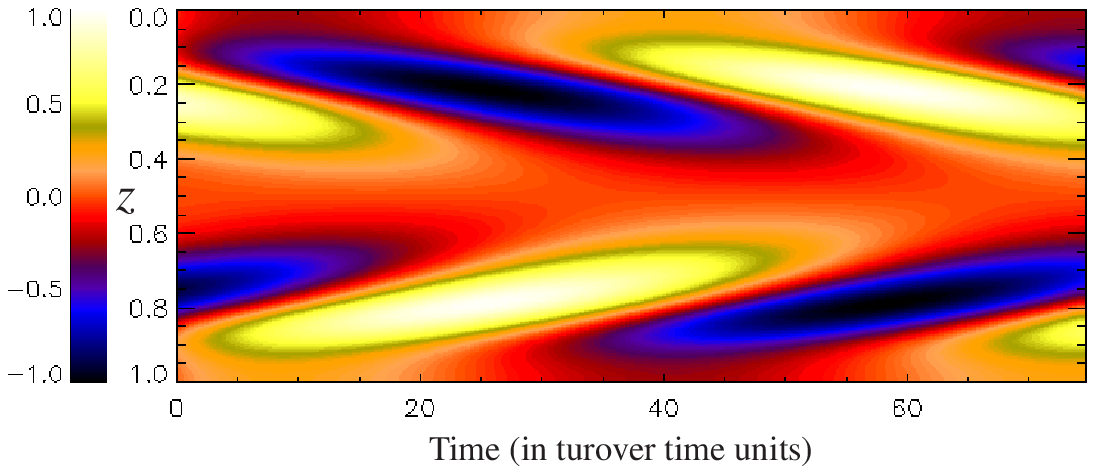}\\
      \includegraphics[width=90mm]{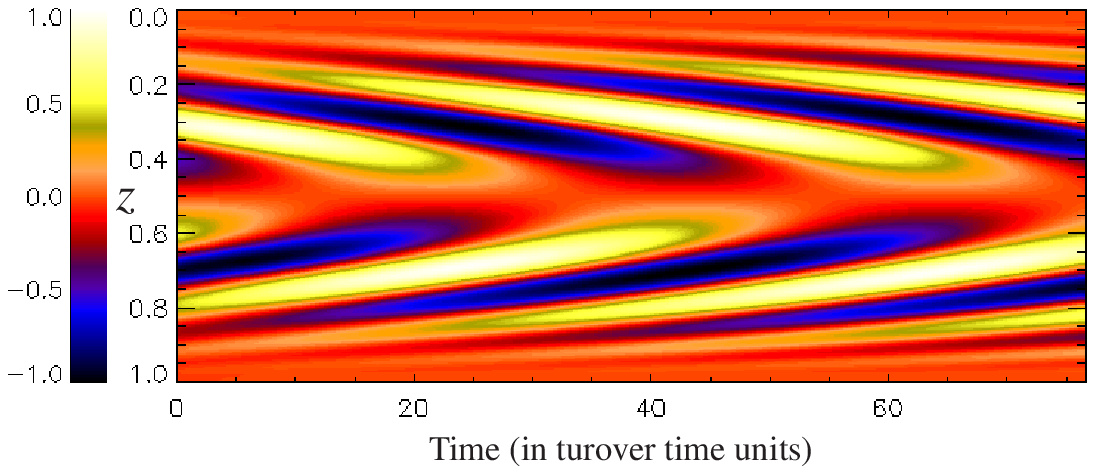}
    \caption{Horizontal average of $B_x$ versus depth and time. The amplitude of the field is compensated to remove the exponential growth and time is scaled in units of the turnover time $\lambda/U_{\textrm{max}}$ where $\lambda$ is the aspect ratio of the domain and $U_{\textrm{max}}$ is the maximum velocity of the flow. The Taylor number is $T=10^8$. The magnetic diffusivity is $\eta=5\times 10^{-3}$ at the top and $\eta=5\times10^{-5}$ at the bottom.\label{fig:butterflys}}
\end{figure}

After some transient phase, the magnetic energy is varying exponentially with time, as expected from the kinematic nature of this problem.
Figure \ref{fig:bx} shows the typical horizontal topology of the magnetic energy after the transient phase, for different $\eta$ and for $T=10^8$.
As the magnetic diffusivity decreases, the field tends to be concentrated at the edges of the convective cells.
Note that the vertical structure of the magnetic field is also becoming more and more complicated as $\eta$ decreases (not shown).
The growth rate of the total magnetic energy is shown on figure \ref{fig:kinmag} as large empty symbols.
For all three Taylor numbers considered here, we observed dynamo action.
For $T=10^8$, dynamo action is first observed for $\eta\approx0.0176$, whereas the critical diffusivity is $\eta\approx0.015$ for $T=10^{10}$ and $\eta\approx0.0115$ for $T=10^{12}$.
Note that the fact that the critical value of the magnetic diffusivity for dynamo action does depend on the Taylor number, and therefore on $\epsilon$, is inconsistent with the mean-field model discussed in section \ref{sec:mean}.
This is probably due to the fact that the values of $\epsilon$ considered here are too large for the mean-field model to be rigorously applicable.
%
\begin{figure*}
      \includegraphics[width=40mm]{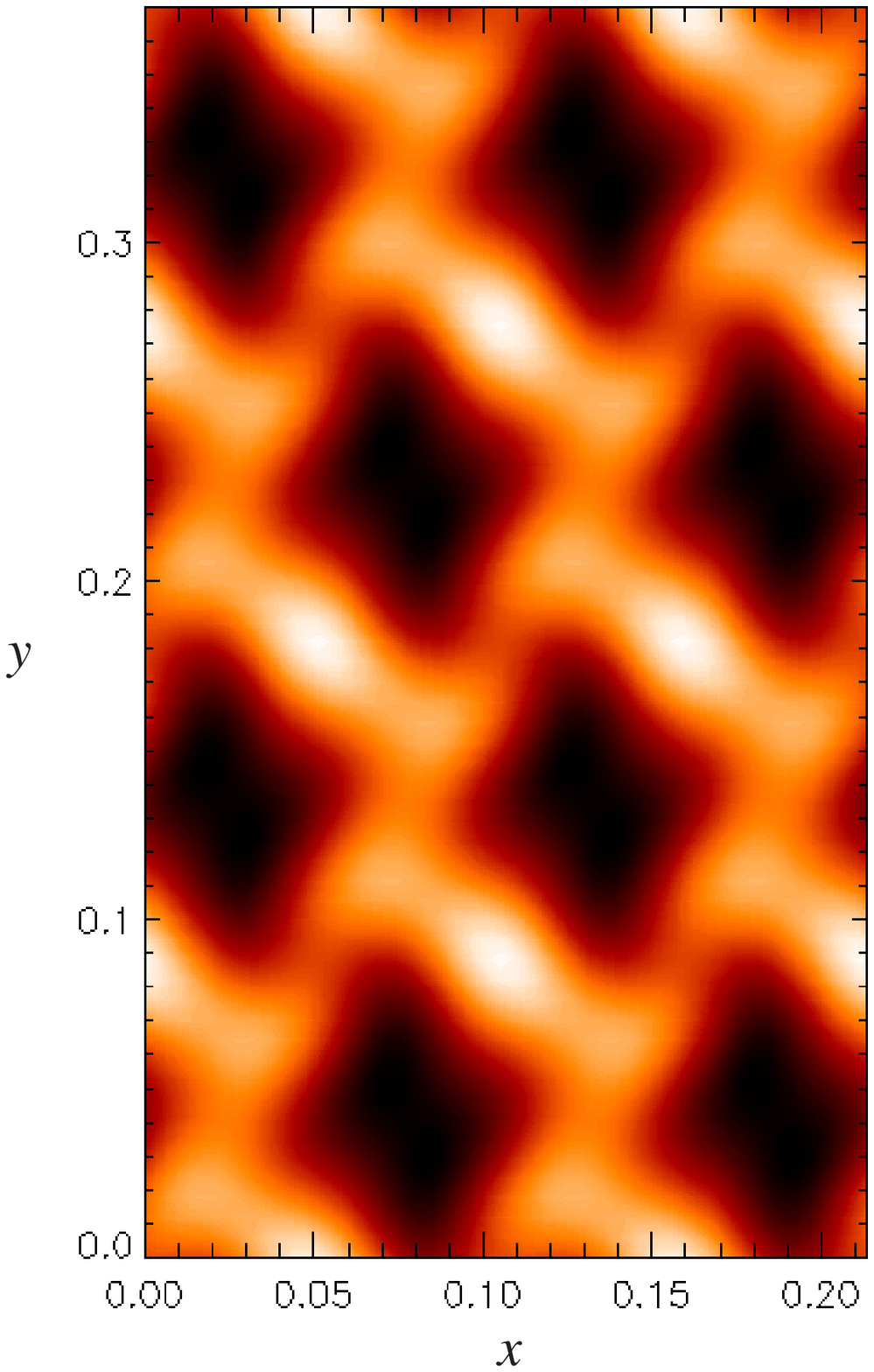}\hspace{10mm}
      \includegraphics[width=40mm]{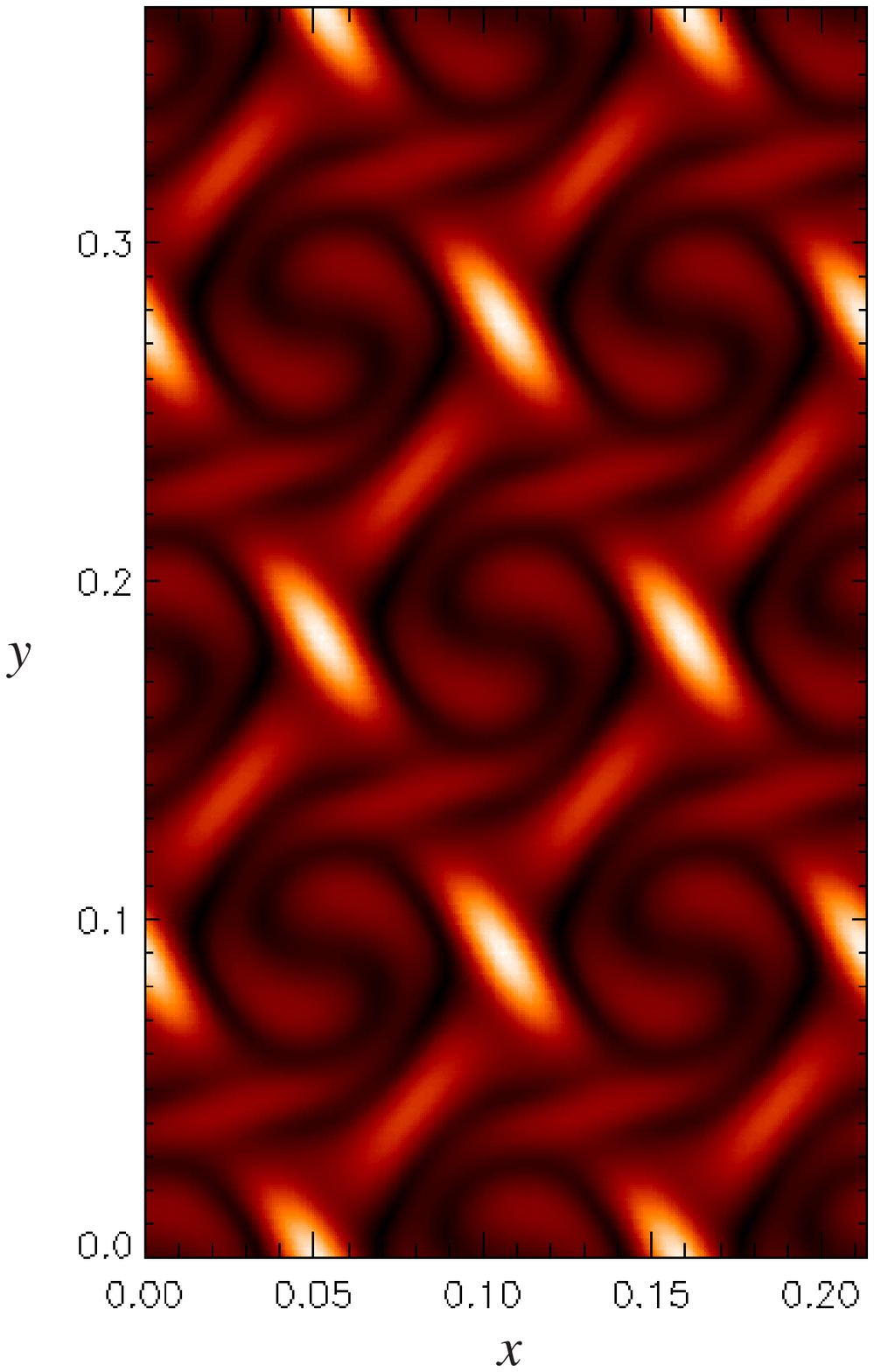}\hspace{10mm}
      \includegraphics[width=40mm]{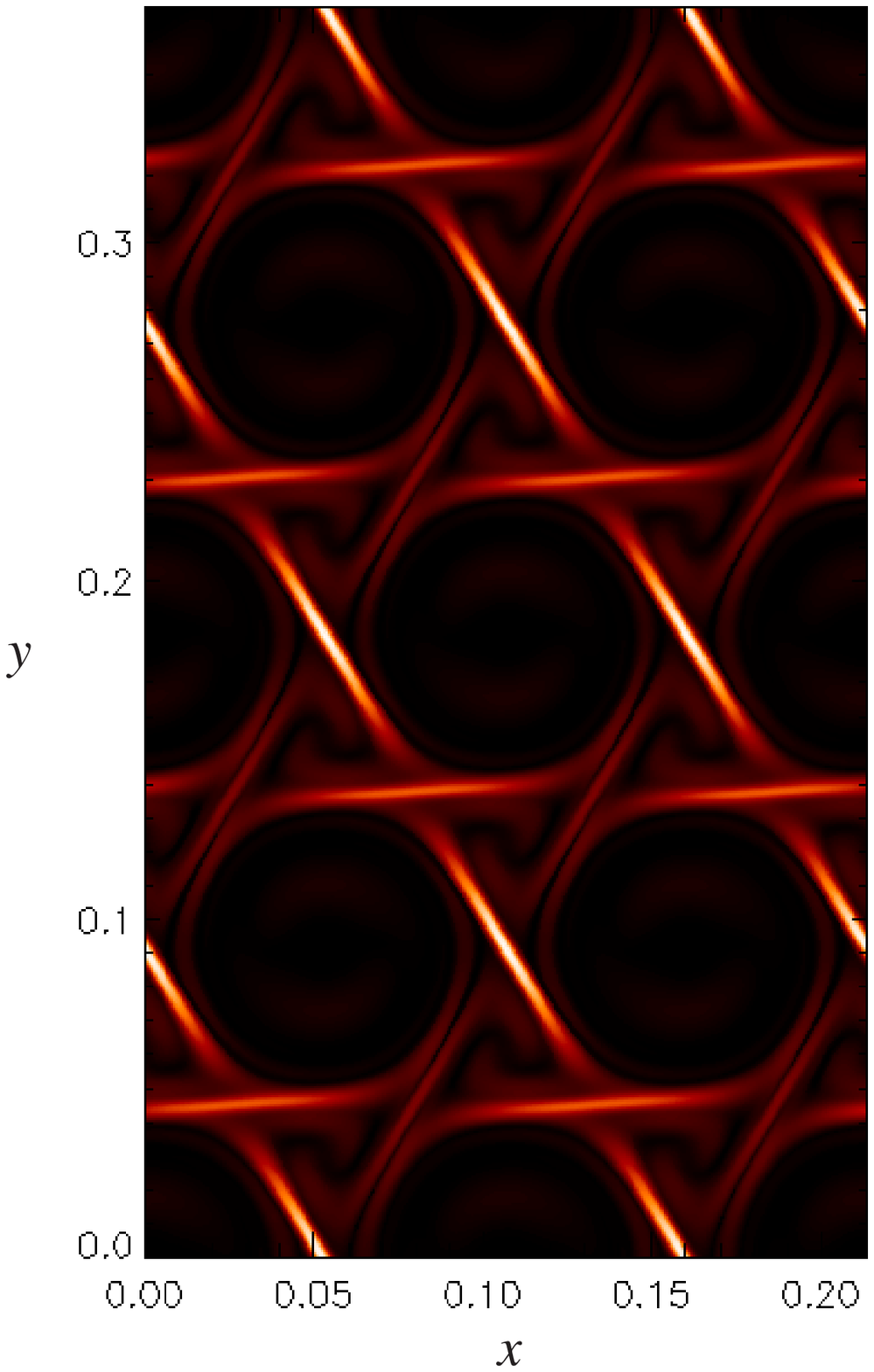}
    \caption{Magnetic energy $\bm{B}^2$ in a horizontal plane located at $z=0.25$. For the sake of clarity, the aspect ratio has been multiplied by two by plotting four copies of the domain side by side. Bright and dark tones correspond to opposite polarity. From left to right, $\eta=10^{-2}$ (close to the critical value for dynamo action), $\eta=10^{-3}$ and $\eta=5\times10^{-5}$ (small-scale dynamo action is marginal). The local orientation of the mean-field is the same in all cases.}
    \label{fig:bxhex}
\end{figure*}
%

As the magnetic diffusivity is decreased, the growth rate increases up to a maximum.
This maximum growth rate  is reached at smaller diffusivities as the Taylor number is increased.
After this point, a further decrease in $\eta$ corresponds to a decrease in the growth rate.
This reduction in the growth rate was not predicted by the mean-field model, since the growth rate was a monotonic function of $\lambda$ in that case.
Note that as the Taylor number is increased, the agreement between the model and the simulations is better, since the range of validity of the mean-field model increases.

Let us now describe the nature of the dynamo action at onset.
Following \cite{cattaneo06} and \cite{favier2012b}, it is helpful to define the quantity
\begin{equation}
\label{eq:gamma}
\Gamma(z)=\frac{\left<\bm{B}\right>^2}{\left<\bm{B}^2\right>} \ ,
\end{equation}
which is the ratio between the magnetic energy contained in the mean-field and the total magnetic energy at a given depth.
Our initial condition corresponds to $\Gamma=0$ everywhere.
For a small-scale dynamo, this ratio is expected to be very small, whereas larger values are expected for a large-scale dynamo.
In all cases, we observed an increase in $\Gamma$ with time, until the system reaches its exponential growth for which $\Gamma$ is steady.
$\Gamma(z)$ is presented on the right part of figure \ref{fig:kinmag}, for the case $T=10^8$.
At onset (\textit{i.e.} for $\eta=0.015$), the maximum value of $\Gamma(z)$ occurs at $z\approx0.69$, which also corresponds to the maximum of the relative kinetic helicity, as shown on figure \ref{fig:hel}.
For $T=10^8$, the maximum value is $\Gamma\approx0.69$, showing the existence of a dominating mean-field across the convective layer.
Note that the magnetic fluctuations are still of the same order as the mean horizontal field.
This is expected since this Taylor number is not large enough for the mean-field model to be applicable at the dynamo onset, as discussed in section \ref{sec:mean}.
As the Taylor number is increased, the maximum value of $\Gamma$ close to onset increases, confirming that the kinematic dynamo is of mean-field type at onset. 
To compare, similar but turbulent flows usually produce much smaller values of $\Gamma$, typically of order $10^{-3}$ (see for example \cite{cattaneo06} and \cite{favier2012b}).
As $\eta$ decreases, we observe a decrease in $\Gamma$, showing that the magnetic field is now dominated by its fluctuating components.
This decrease in $\Gamma$ as the magnetic diffusivity decreases is observed for all three Taylor numbers considered here.

The structure and evolution of this dominant horizontal mean magnetic field is shown on figure \ref{fig:butterflys} for $T=10^8$.
The horizontal average of $B_x$ is plotted versus depth and time in a ``butterfly'' diagram for two different magnetic diffusivities.
The result is normalised in order to compensate for the exponential growth.
At any given time, the structure of the eigenmode predicted by the mean-field model discussed in section \ref{sec:mean} is qualitatively recovered (see figure \ref{fig:evps} for $\lambda=12.2$ and $\lambda=20.1$).
As the magnetic diffusivity $\eta$ decreases, smaller wavelengths are observed in the eigenfunction and the frequency of rotation of the mean-field decreases, in accordance with the mean-field model.
Note that for all $\eta$, the mean horizontal field is drifting from the boundaries towards the mid-layer and is anti-symmetric with respect to the mid-layer.
At a fixed depth, this corresponds to a mean horizontal field rotating around the vertical axis with a given frequency.
The sign of this rotation changes between the upper and lower half of the domain.
Note that this type of solution is reminiscent of the dynamo solution for the Roberts flow \cite{roberts72} (but since the helicity is changing sign across the mid-layer, so is the rotation rate of the rotating horizontal mean-field), and also shares some similarities with the numerical solution of \cite{stell04} which reported a strong horizontal mean-field rotating with time.
This is not surprising as the Roberts flow is locally very similar to the flow discussed in this section, and by Stellmach \& Hansen \cite{stell04} who considered rotating Boussinesq convection at very low Ekman numbers (\textit{i.e.} very large Taylor numbers) close to the onset of convection.

To further investigate the nature of the dynamo action at onset, we run an additional set of simulations, identical to the previous ones in most respects.
However, instead of solving the mean induction equation 
\begin{equation}
\label{eq:meanfield}
\frac{\partial \left<\bm{B}\right>}{\partial t}=\bm{e}_z\times\frac{\partial \left<\bm{\mathcal{E}}\right>}{\partial z}+\eta\frac{\partial^2 \left<\bm{B}\right>}{\partial z^2} \ ,
\end{equation}
for the horizontally-averaged magnetic field, where $\bm{\mathcal{E}}=\left<\bm{u}\times\bm{B}\right>$ is the horizontally averaged electromotive force, we \textit{artificially} constrain these mean magnetic fields to be zero everywhere in the layer.
This approach has already been used by \cite{ponty2011} when studying the transition between large-scale and small-scale dynamos in the Roberts flow, and by \cite{favier2012b} to measure the $\alpha$-effect generated by rotating turbulent convection.
In these artificial simulations, the only possible dynamo is of small-scale nature as the mean electromotive force is neglected.
We show on figure \ref{fig:kinmag}(a) the growth rates of such simulations as small full symbols.
The onset for dynamo action in this case corresponds to $\eta\approx 2.8 \times 10^{-4}$ for $T=10^8$ and $\eta\approx 1.3 \times 10^{-4}$ for $T=10^{10}$.
Note that the critical magnetic Reynolds number based on the horizontal scale of motion is now the same for both Taylor numbers.
This further confirms that the previous dynamo action observed for smaller values of $\eta$ is of mean-field type, since small-scale dynamo is not possible for that range of parameters.
However, as $\eta$ decreases, small-scale dynamo eventually becomes possible and the growth rate of the dynamo then decreases, along with $\Gamma$.
This transition between large-scale dynamo and small-scale dynamo action shares some similarities with what has been observed for the Roberts flow \cite{ponty2011} and for helically forced flows in spherical shells \cite{richardson2012}.
%
%
\section{Hexagonal pattern \label{sec:hexa}}
\begin{figure}
      \includegraphics[width=80mm]{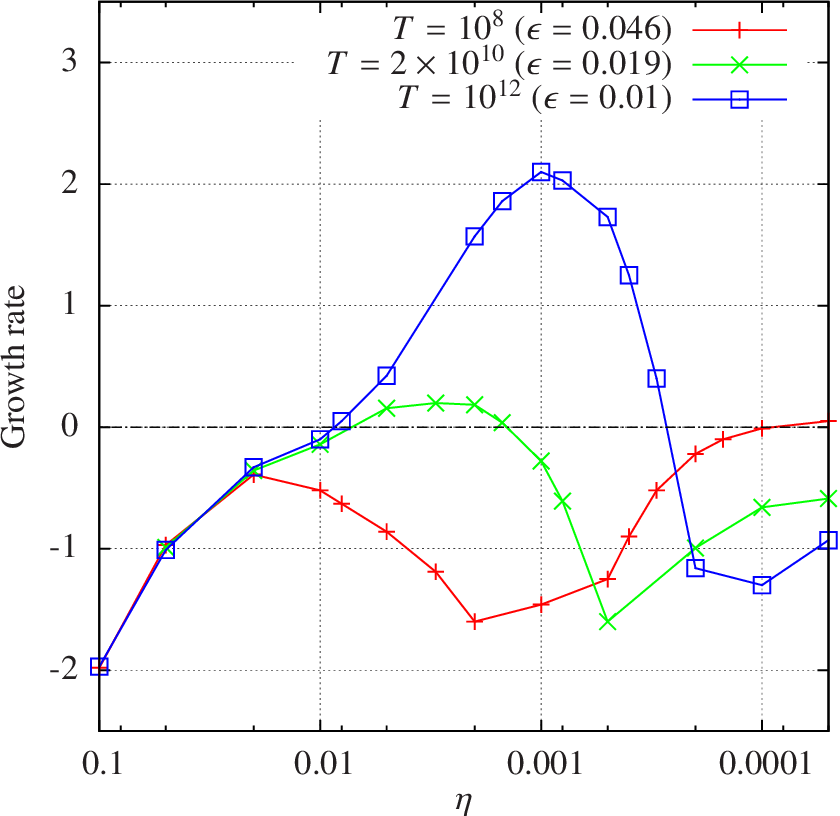}
    \caption{Growth rate of the magnetic energy versus $\lambda$ for three different Taylor numbers in the hexagonal pattern case. For $T=10^8$, all the growth rates are negative apart from the smallest value of $\eta$ which is very close to criticality but the growth rate is positive.\label{fig:grhex}}
\end{figure}

We now consider the hexagonal cells.
The flow is defined by equations \eqref{eq:hex1}-\eqref{eq:hex3}.
We use a similar approach to the one used in section \ref{sec:square}: the induction equation \eqref{eq:induction} is numerically solved in three dimensions as described in section \ref{sec:model}.
We choose three representative Taylor numbers: $T=10^{8}$, $T=2\times 10^{10}$ and $T=10^{12}$.
The corresponding parameters can be found in Table~\ref{tab:one}.
The most unstable wave number of the convective motion $a$, as defined by equation \eqref{eq:a}, varies between these simulations and the aspect ratios are adjusted accordingly.
As for the simulations in the square pattern, the magnetic field is initialised with a small perturbation with zero net flux.

We show on figure \ref{fig:bxhex} the magnetic energy in a horizontal plane located at $z=0.25$, for different values of the magnetic diffusivity $\eta$.
As for the square pattern, the magnetic field tends to be expelled from the center of the convective cells.
According to the previous mean-field model described in section \ref{sec:mean}, the case $T=10^8$ (for which $\epsilon=0.046$) is not able to sustain a large-scale dynamo.
On figure \ref{fig:grhex}, we present the growth rate of the magnetic energy versus the magnetic diffusivity.
For $T=10^8$, we indeed observe an increase in the growth rate up to $\eta\approx0.02$ followed by a decrease.
A small positive growth rate is obtained at a much smaller diffusivity of $\eta=5\times10^{-5}$.
This dynamo is of small-scale nature and is characterised by $\Gamma\approx10^{-3}$.
Of course, the possibility of small-scale dynamo action is not predicted by the mean-field model, but the lack of a large-scale dynamo is however consistent with the prediction of the mean-field model.
\begin{figure}
      \includegraphics[width=90mm]{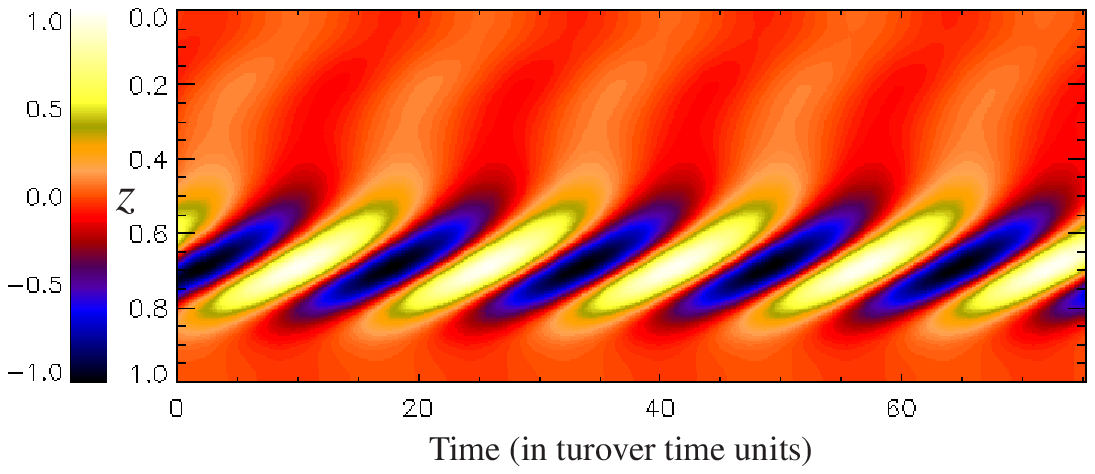}\\
      \includegraphics[width=90mm]{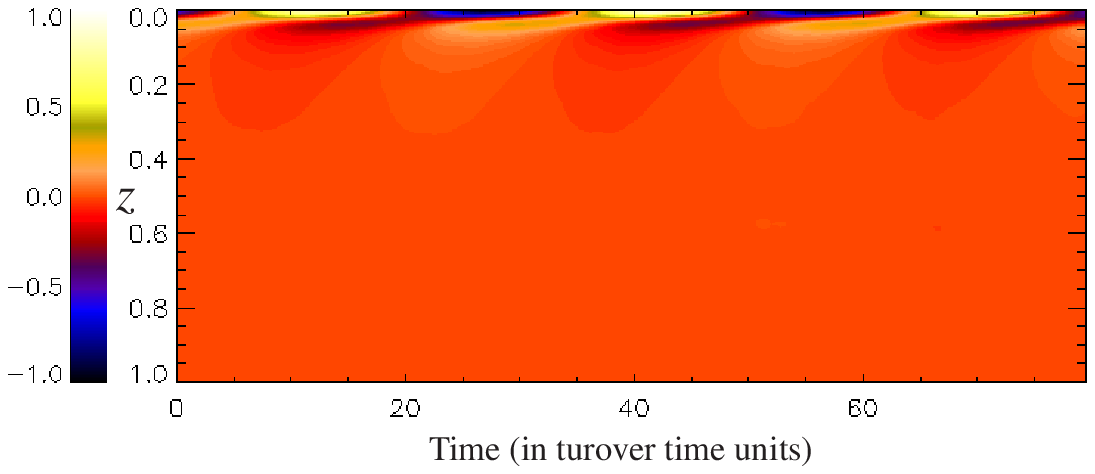}
    \caption{Horizontal average of $B_x$ versus depth and time in the hexagonal case. The amplitude of the field is compensated to remove the exponential growth and time is scaled in units of the turnover time $\lambda/U_{\textrm{max}}$ where $\lambda$ is the aspect ratio of the domain and $U_{\textrm{max}}$ is the maximum velocity of the flow. The Taylor number is $T=2\times10^{10}$ and the magnetic diffusivity is $\eta=5\times 10^{-3}$ at the top (dynamo action is observed) and $\eta=3\times10^{-4}$ at the bottom (no dynamo action).\label{fig:butterfly}}
\end{figure}

For this Taylor number, the hexagonal pattern is not capable of sustaining a large-scale dynamo due to the dominating effect of the pumping velocity.
For $T=2\times10^{10}$, there is a range of parameters for which mean-field dynamo is possible, namely $0.02<\eta<0.07$.
As predicted by the mean-field model, for $\eta>0.07$, the mean-field dynamo is shut down due to an increasingly efficient pumping velocity.
Small-scale dynamo action might be possible at smaller values of the diffusivity, although we could not check this numerically due to resolution constraints.
Similar results are obtained for $T=10^{12}$, but the range of magnetic diffusivity for which a large-scale dynamo is observed increases, as the pumping velocity is only affecting the dynamo at smaller value of $\eta$.
Ultimately, for an infinite Taylor number, the pumping velocity will always be negligible, so that the hexagonal pattern would be qualitatively similar to the square pattern, as predicted by the mean-field model.
Note that the reduction in the dynamo growth rate observed for the square pattern is not predicted by the mean-field model and is related to small-scale dynamo action, whereas the case of the hexagonal pattern and the eventual disappearance of the large-scale dynamo is fully explained by the mean-field model.

The structure of the eigenmodes is shown as a space-time diagram on figure \ref{fig:butterfly}.
We observe a qualitatively similar behaviour as in the mean-field model.
When dynamo action is possible, the mean-field is dominant in the lower half of the domain, as already observed on figure \ref{fig:evph}.
At smaller magnetic diffusivity, the mean-field is dominantly advected towards the upper boundary (the pumping velocity is directed upward) and the dynamo is effectively shut down.

%
\section{Effect of changing the boundary conditions \label{sec:bou}}

\begin{figure}
      \includegraphics[width=80mm]{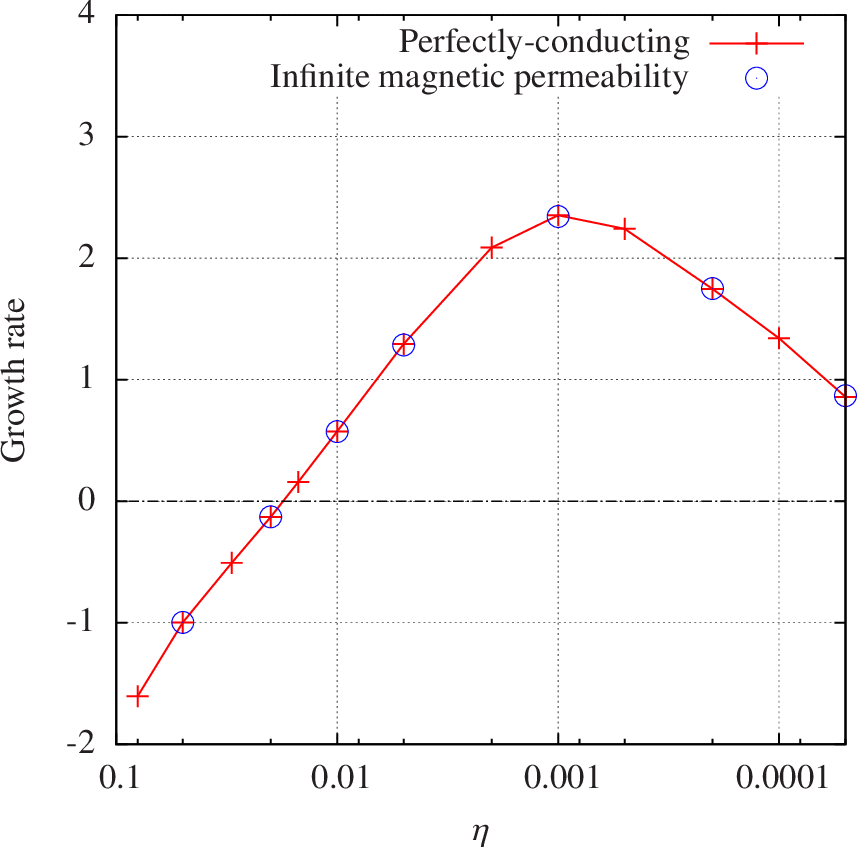}\\
      \includegraphics[width=80mm]{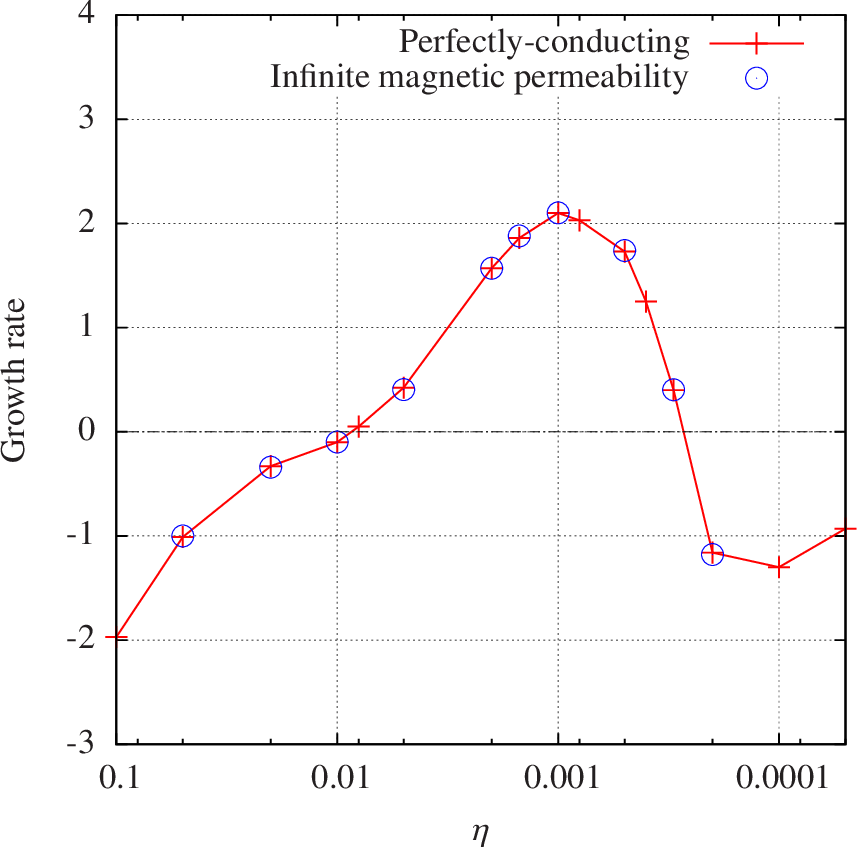}
    \caption{Growth rate of the magnetic energy versus $\eta$. Left: results corresponding to the square pattern at $T=10^8$. Right: results corresponding to the hexagonal pattern at $T=10^{12}$. In each case, we compare the two types of boundary conditions, perfectly-conducting (\textit{i.e.} $B_{x,z}=B_{y,z}=B_z=0$) and vertical (\textit{i.e.} $B_x=B_y=B_{z,z}=0$).\label{fig:changebcs}}
\end{figure}
Finally, we look at the effect of changing the magnetic boundary condition.
All the previous results have been obtained for perfectly-conducting boundaries for which the magnetic field lines are horizontal at $z=0$ and $z=1$.
Instead, we now assume that the boundaries have an infinite magnetic permeability so that the magnetic field lines reconnect perpendicularly to the boundaries.
The magnetic field tangent to the boundaries vanishes along with the normal current density and we therefore have $B_x=B_y=\partial_zB_z=0$ at $z=0$ and $z=1$.
Of course we now have to abandon the constraint of zero horizontal magnetic flux (see equation \eqref{eq:consflux}), since this is not a conserved quantity using this new set of boundary conditions.

Although the magnetic field topology is now very different, the growth rate of the magnetic energy is exactly the same as in the perfectly-conducting case, for all the values of $\eta$ considered here, and for both the mean-field equations and the full three-dimensional simulations.
This remarkable result is in fact general and not a specificity of the current model.
It is due to the adjointness property of the induction equation as discussed by \cite{gibrob67,proctor77a,proctor77b}.
Provided one can reverse the direction of the flow by an appropriate set of transformations which leave the boundaries invariant (we call this class of flows reversible flows in the following), the growth rate of any kinematic dynamos will be exactly the same whether the boundaries are made of a perfect electrical conductor or have an infinite magnetic permeability.
A formal demonstration and additional examples of this rigorous result can be found in \cite{favierproctor}.
\begin{figure}
      \includegraphics[width=90mm]{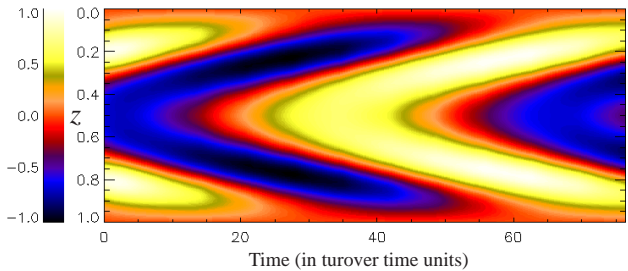}
    \caption{Horizontal average of $B_x$ versus depth and time in the square pattern at $T=10^8$ and $\eta=5\times 10^{-3}$. The upper and lower boundaries have an infinite magnetic permeability so that magnetic field lines are normal to them. The amplitude of the field is normalised to compensate for the exponential growth and time is scaled in units of the turnover time.\label{fig:butterflyv}}
\end{figure}

The particular flows considered in this paper are all reversible so that this general result is applicable in this case.
In the case of the square pattern, a simple horizontal translation can change $\bm{u}$ into $-\bm{u}$.
For the hexagonal pattern, the direction of the flow is reversed under point reflection with respect to the mid-layer, effectively changing the sign of all three spatial coordinates.
Using the mean-field model described in section \ref{sec:mean} but imposing a vanishing horizontal magnetic field at $z=0$ and $z=1$ (the conditions \eqref{eq:consflux} are also relaxed since they are only valid in the perfectly-conducting case), the exact same growth rates as in the perfectly-conducting case are obtained.
The associated eigenfunctions are however different: they are now symmetric with respect to the mid-layer.

In addition, a comparison of the growth rates obtained with the direct numerical simulations described in section \ref{sec:model} and for the two different set of boundary conditions is shown in figure \ref{fig:changebcs}.
We compare the magnetic energy growth rates for the square pattern at $T=10^8$ and for the hexagonal pattern at $T=10^{12}$.
For all diffusivities considered here, the growth rates of the magnetic energy is indeed the same for the two types of magnetic boundary condition.
Interestingly enough, in the case of the square pattern and with vertical boundary conditions, we obtain a similar space-time diagram as on figure \ref{fig:butterflys}, but the horizontal mean-field is rotating in the opposite direction (see figure \ref{fig:butterflyv}), so that each horizontal component is drifting from the mid-layer towards the boundaries.
The eigenfunction is also symmetric with respect to the mid-layer whereas it was anti-symmetric in the case of perfectly-conducting boundary conditions, which is consistent with the results of the mean-field model (not shown).
Although the two eigenfunctions are qualitatively different, the real parts of the associated eigenvalues are rigorously the same for any magnetic diffusivity.
This result holds for both large- and small-scale dynamos since it is a property of the general induction equation for reversible flows.
%
%
\section{Conclusion}
In this paper, we investigated the kinematic dynamo action in rotating convective flows.
By considering the onset of convection, we were able to select between square and hexagonal patterns.
The flow is then analytically prescribed and the induction equation is solved numerically.
We first use a reduced model based on a mean-field approach and we then consider direct numerical simulations of the full induction equation.
For the square pattern, we observe first a dynamo of mean-field type where the mean electromotive force compensates for diffusion.
As the magnetic Reynolds number increases, a transition towards small-scale dynamo action is found and the magnetic energy is dominated by its fluctuating component.
This transition also corresponds to a decrease in the kinematic growth rate of the magnetic energy.

For the hexagonal pattern, the situation is more complicated.
Due to the asymmetry between the up and down flows, an effective pumping velocity appears.
This effect corresponds to the off-diagonal terms in the classical $\alpha$-tensor of mean-field electrodynamics.
While such an effect can be removed in any vertically-invariant flows (such as the Roberts flow), it is not the case in our plane-layer confined model.
It ensues that the previously observed mean-field dynamo can disappear if the Taylor number is too small (while being large enough to justify the small $\epsilon$ regime required by the mean-field model).
This is a surprising example where some of the terms of the $\alpha$ tensor are actually be unhelpful for dynamo action.
For sufficiently large Taylor number, the pumping velocity becomes negligible and the results become qualitatively similar to the square pattern: a large-scale dynamo is observed at onset whereas small-scale dynamo action is probably possible at much larger magnetic Reynolds numbers (although we only numerically observe a small-scale dynamo for the case $T=10^8$, where no large-scale dynamo is possible).
These conclusions are derived using both a mean-field model and direct numerical simulations of the full induction equation with a prescribed velocity field.
Note that the fact that the pumping velocity decreases with the Taylor number is consistent with the prediction of \cite{tobias2001}, although they considered the fully-turbulent regime in their case.

Many aspects of this problem remain to be studied.
It is worth mentioning that we focus in this paper on dynamo solutions having the same periodicity as the background flow.
We considered numerical simulation with larger aspect ratios in order to allow for spatially-modulated solutions, but we found none that grew more rapidly.
A more complete analysis is however required in the general case, as we cannot exclude the possibility of more efficient sub-harmonic dynamos at even larger aspect ratios.
In addition, the nonlinear saturation of these dynamos is of interest.
This would require the solution of the momentum equation coupled with the induction equation.
Self-consistently obtaining these square and hexagonal flows can be challenging in itself.
We managed to produce both square and hexagonal patterns in rapidly-rotating weakly stratified compressible convection, by varying the horizontal aspect ratio of the numerical domain.
The kinematic dynamo properties of such flows are very similar to what has been described in the present paper, even if the flows significantly depart from equations \eqref{eq:square1}-\eqref{eq:square3} and \eqref{eq:hex1}-\eqref{eq:hex3}.
Much remains to be done concerning the nonlinear saturation of these kinematic dynamos.

As mentioned in the introduction, the turbulent regime far from onset is still problematic when it comes to its large-scale dynamo action capability.
We showed that both squares and hexagons are capable of sustaining a large-scale dynamo, providing the Taylor number is large enough.
An interesting question is how does this dynamo solution behaves as the flow becomes decorrelated in space and time.
It is known that the mean induction is dramatically reduced as the flow becomes less spatially correlated \cite{courvoisier2009}, and it would therefore be interesting to study how these well-defined mean-field dynamos behave when the Rayleigh number is increased, introducing spatio-temporal chaos in the flow.
This undergoing study should fill the gap between mean-field dynamos in simple analytic flows and turbulent dynamos, where the distinction between small-scale and large-scale dynamos is often unclear.

\bibliographystyle{plain}
\bibliography{biblio}

\end{document}